\documentclass[twocolumn, times]{aastex631}

\usepackage{amsmath, latexsym, color, verbatim}

\shorttitle{MIRI MRS Flux Calibration}
\shortauthors{Law et al. 2024}

\newcommand{\nevi}{\textrm{[Ne\,{\sc vi}]}}
\newcommand{\nev}{\textrm{[Ne\,{\sc v}]}}

\newcommand{\ofour}{\textrm{[O\,{\sc iv}]}}

\begin{document}

\title{The James Webb Space Telescope Absolute Flux Calibration. III. Mid-Infrared Instrument Medium Resolution IFU Spectrometer}

\author[0000-0002-9402-186X]{David R.\ Law}
\affiliation{Space Telescope Science Institute, 3700 San Martin Drive, Baltimore, MD, 21218, USA}

\author[0000-0003-2820-1077]{Ioannis Argyriou}
\affiliation{Institute of Astronomy, KU Leuven, Celestijnenlaan 200D, 3001 Leuven, Belgium}

\author[0000-0001-5340-6774]{Karl D.\ Gordon}
\affiliation{Space Telescope Science Institute, 3700 San Martin Drive, Baltimore, MD, 21218, USA}

\author[0000-0003-4520-1044]{G.\ C.\ Sloan}
\affiliation{Space Telescope Science Institute, 3700 San Martin Drive, Baltimore, MD, 21218, USA}
\affiliation{Department of Physics and Astronomy, University of North Carolina, Chapel Hill, NC 27599-3255, USA}

\author[0000-0002-1257-7742]{Danny Gasman}
\affiliation{Institute of Astronomy, KU Leuven, Celestijnenlaan 200D, 3001 Leuven, Belgium}

\author[0000-0002-2041-2462]{Alistair Glasse}
\affiliation{UK Astronomy Technology Centre, Royal Observatory, Edinburgh, Blackford Hill, Edinburgh EH9 3HJ, United Kingdom}

\author[0000-0003-3917-6460]{Kirsten Larson}
\affiliation{Space Telescope Science Institute, 3700 San Martin Drive, Baltimore, MD, 21218, USA}

\author[0000-0001-5834-9588]{Leigh N.\ Fletcher}
\affiliation{School of Physics and Astronomy, University of Leicester, University Road, Leicester, LE1 7RH, UK}

\author[0000-0002-0690-8824]{Alvaro Labiano}
\affiliation{Telespazio UK for the European Space Agency, ESAC, Camino Bajo del Castillo s/n, 28692 Villanueva de la Ca\~nada, Spain}
\affiliation{Centro de Astrobiolog\'ia (CAB), CSIC-INTA, ESAC, Carretera de Ajalvir km4, 28850 Torrej\'on de Ardoz, Madrid, Spain.}

\author[0000-0002-6296-8960]{Alberto Noriega-Crespo}
\affiliation{Space Telescope Science Institute, 3700 San Martin Drive, Baltimore, MD, 21218, USA}

\begin{abstract}
We describe the spectrophotometric calibration of the Mid-Infrared Instrument's (MIRI) Medium Resolution Spectrometer (MRS) aboard the James Webb Space Telescope (JWST).
This calibration is complicated by a time-dependent evolution in the effective throughput of the MRS; this evolution is strongest at long wavelengths, approximately a factor of 2 at 25 $\micron$ over the first two years of the mission.  We model and correct for this evolution through regular observations of internal calibration lamps.
Pixel flatfields are constructed from observations of the infrared-bright planetary nebula NGC 7027, and photometric aperture corrections from a combination of theoretical models and observations of bright standard stars.
We tie the 5--18 $\micron$ flux calibration to high signal/noise (S/N; $\sim 600-1000$) observations of the O9 V star 10 Lacertae, scaled to the average calibration factor of nine other spectrophotometric standards.
We calibrate the 18--28 $\micron$ spectral range using a combination of observations of main belt asteroid 515 Athalia and the circumstellar disk around young stellar object SAO 206462.
The photometric repeatability is stable to better than 1\% in the wavelength range 5--18 $\micron$, and the S/N ratio of the delivered spectra is consistent between bootstrapped measurements, pipeline estimates, and theoretical predictions.
The MRS point-source calibration agrees with that of the MIRI imager to within 1\% from 7 to 21 $\micron$ and is approximately 1\% fainter than prior Spitzer observations, while the extended source calibration agrees well with prior Cassini/CIRS and Voyager/IRIS observations.

\end{abstract}



\section{Introduction}

Spectrophotometric (i.e., ``flux'') calibration is an essential part of the calibration
of any astronomical spectrometer.  Accurate relative flux calibration is essential both
to reliably determine the shape of a given spectrum, and for determining line ratios
that encode fundamental parameters (temperature, metallicity, density, etc) of the object in question.
There are few truly absolute calibration sources however, and in practice calibrations
are typically bootstrapped from comparing observations of spectrophotometric standard
stars to high-resolution models built upon extensive multi-wavelength data.
In the mid-infrared however this can be challenging, both because models are more
uncertain, and also because even the brightest standard stars become quite faint
longward of 20 microns when dispersed at moderate spectral resolution.

In this contribution, we discuss the 
calibration of the MIRI \citep{rieke15,wright23} Medium Resolution Spectrometer \citep[MRS;][]{wells15,argyriou23} aboard the James Webb Space Telescope \citep[JWST;][]{gardner23}.  The MRS consists of four discrete and roughly concentric integral field units (IFUs) that jointly cover the wavelength range 5--28 $\micron$ using a series of dichroic beam splitters.  Each of these four channels is split into thirds using a pair of wheels hosting the dichroic and grating assemblies (DGA); the three DGA wheel positions respectively select light in the first (A/SHORT), middle (B/MEDIUM), and last (C/LONG) third of light within the wavelength range covered by each of
the IFU wavelength channels.  Light from channels 1/2 and 3/4 are dispersed onto the IFUSHORT and IFULONG detectors respectively.  Twelve spectral bands are thus required in order to fully cover the wavelength range 5--28 $\micron$.

Calibration of the MRS is made challenging by a few key factors that must be accounted for in our analysis.  First, even standard stars of naked-eye magnitudes (V = 5) are faint compared to the detector noise and thermal background at the longest wavelengths \citep{rigby23}.
Second, observations have determined that the effective throughput of the MRS varies with time, changing by more than 50\% at long wavelengths over the first two years of the mission.
The MRS also experiences significant (tens of percent) spectral fringing from constructive and destructive interference within the MIRI detectors and other optical components; the magnitude and phase of this fringing varies rapidly with source geometry and position \citep[see, e.g.,][]{argyriou20}.  
Finally, charge migration (also known as the brighter-fatter effect; BFE) results in charge-spilling that can bias adjacent pixel ramps
\citep[][]{argyriou23b,gasman24}.

We present here the spectrophotometric calibration of the MIRI MRS in the standard JWST calibration pipeline made available by STScI.  This effort incorporates calibration data obtained over the first 2.5 years of science observations and builds upon the 
commissioning and ground-based calibration methodology described by 
\citet{argyriou23}.  Unlike these early calibrations however, sufficient data now exist to fully separate the 2D flatfield and 1D photometric response terms that were previously combined by necessity.
This approach is designed to provide a general-purpose calibration that can be used for all MRS observations regardless of the astronomical target, and is thus distinct from efforts to optimize both the fringe and photometric calibration for specific point-source data \citep[e.g.,][]{gasman23}.

In Section \ref{flats.sec} we describe our approach to flatfielding the MRS data, combining both static models of the spectral fringing with traditional pixel flats derived via self-calibration from observations of bright nebulae.  In Section \ref{tputloss.sec}
we characterize the MRS time-dependent throughput loss, and develop a mathematical model
for the loss fraction as a function of time since launch.  We discuss the adopted aperture
corrections based on our latest understanding of the MRS 
point-spread function in
Section \ref{apcorr.sec}.  In Section \ref{photom123.sec} we discuss our approach to calibrating the spectrophotometric response in MRS Channels 1-3 ($\lambda < 18\micron$) using observations of bright spectrophotometric calibration standard stars.  These stars are too faint to derive reliable calibrations at longer wavelengths, and in Section \ref{photom4.sec} we describe our hybrid approach to flux calibration for MRS Channel 4 (i.e., $\lambda > 18\micron$) using observations of bright red targets.  We assess the repeatability of our calibration in Section \ref{repeat.sec} and discuss cross-calibration against other JWST instruments and missions in Section \ref{discussion.sec}.

Throughout this work we use Build 11.0 of the JWST calibration pipeline \citep{bushouse22}, corresponding to jwst software tag 1.15.1 and CRDS reference file context 1276.

\section{Flatfields}
\label{flats.sec}

As discussed by \citet{rieke15} and \citet{ressler15}, the MIRI MRS uses a pair of 1024$\times$1024 arsenic-doped silicon
impurity band conduction detector arrays to record the light dispersed by the spectrometer optics.  The relative response of individual pixels in these arrays must be calibrated prior to deriving a single 1-D wavelength-dependent photometric calibration vector for each band.  In the JWST science calibration pipeline \citep{bushouse22} these calibration flatfields have two components: a static fringe flat representing the periodic amplitude modulations resulting from interference within the instrument for a uniformly illuminated extended source, and a pixel flat representing the relative area of each pixel on the sky convolved with the individual pixel responsivity.

\begin{figure}[!]
\epsscale{1.1}
\plotone{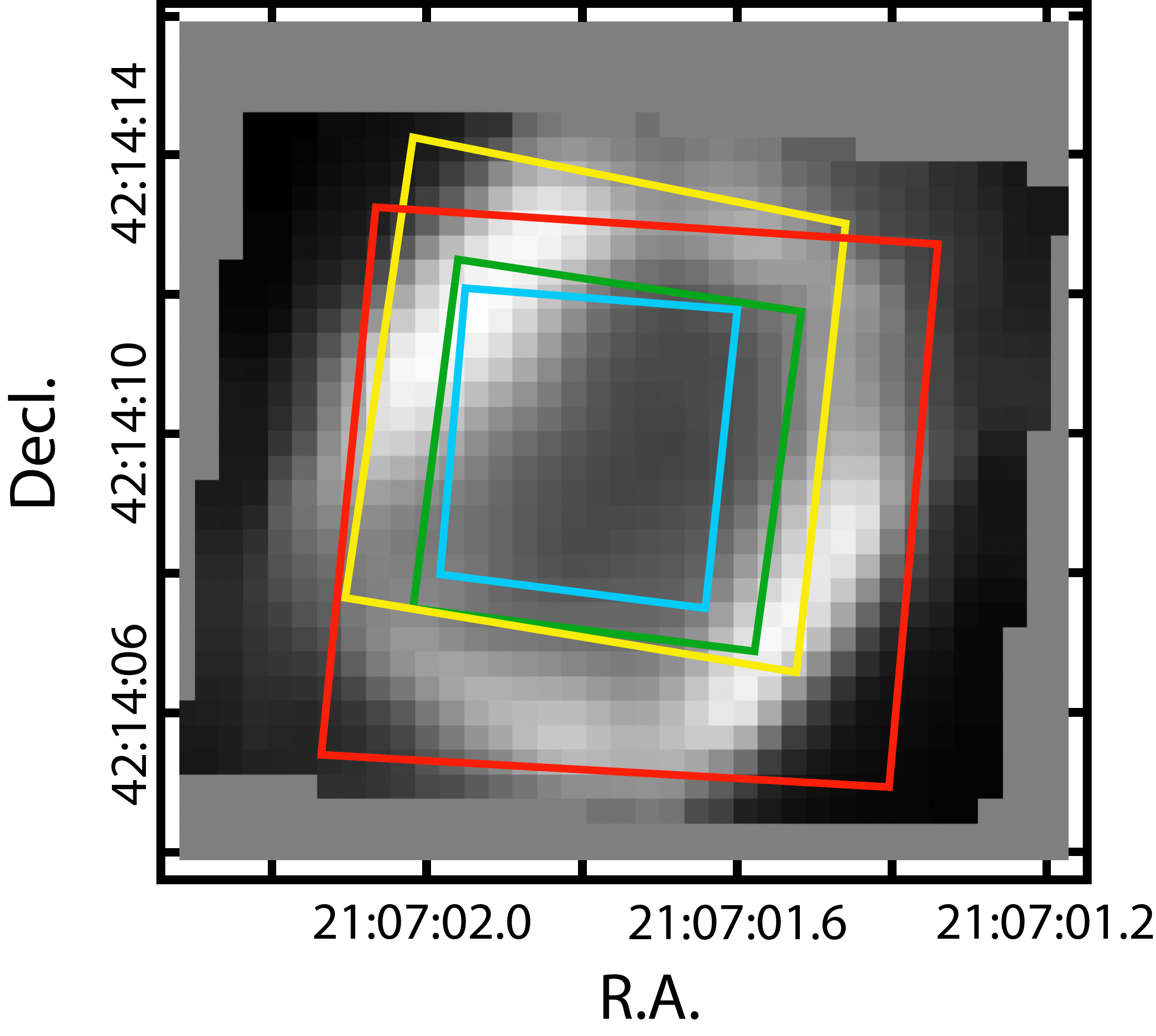}
\caption{18.5 $\mu$m image of NGC 7027 (greyscale) created from a 3$\times$3 mosaic of MRS Channel 4A observations from PID 1523.  The blue, green, yellow, and red rectangles represent the single-tile MRS fields of view for channels 1, 2, 3, and 4 respectively.
}
\label{ngc7027.fig}
\end{figure}

Both ``fringe flats'' and ``pixel flats'' have been derived via self-calibration from observations of the bright planetary nebula NGC 7027 (Figure \ref{ngc7027.fig}).  This source was observed in JWST observing Cycles 1 and 2 (PID 1523 and 4489 respectively) using a 3$\times$3 mosaic and a 2-point dither pattern optimized for extended sources at each tile location, providing a total of 18 individual pointings per Cycle across the nebula.  A similar set of observations of planetary nebula NGC 6543 was obtained in commissioning \citep[PID 1031, 1047; see][]{argyriou23} and used to derive the initial commissioning fringe flats, although the S/N of these data was typically not as high as the NGC 7027 observations.


\subsection{Fringe Flats}
\label{fringeflat.sec}

The process of deriving static fringe flats from observations of NGC 7027 is described
in detail by \citet{crouzet24}.  These fringe flats characterize the 
interferometric fringing
of the MRS detectors for uniform, spatially extended illumination
in each of the twelve spectral bands.

These flats will thus never be perfectly representative for any astronomical scene, as deviations from uniform illumination will result in a slightly different fringing pattern \citep{argyriou20}.  In practice though they are sufficiently close for most extended sources that their application significantly reduces the amplitude of fringing artifacts.  Point sources, however, have a sufficiently different fringing pattern that alternative methods are required to satisfactorily remove them from the data (see discussion in \S \ref{photom123.sec}).


\subsection{Pixel Flats}
\label{pixflat.sec}


While the fringe flats correct for periodic amplitude modulations due to reflections of infrared light within
the detectors, the pixel flats correct for variations in the projected detector pixel area and responsivity at fixed wavelength.
These pixel flats are derived via self-calibration in an iterative process using our 18 different pointings within NGC 7027.

Since the JWST data is calibrated to units of surface brightness (MJy/sr), the largest single term in the pixel flatfield is the relative solid angle of each pixel projected onto the sky.  Our first-pass pixel flat is thus derived by using the MRS distortion solution \citep{patapis24} to compute the footprint of each pixel in square arcseconds, and normalizing by the median pixel footprint in a given band.  This gives flatfields in which effective pixel responsivities range roughly $\pm 10$\% about unity, with strong systematic trends from slice to slice and within slices following the distortion solution.  We then iterate upon this solution in order to both refine the nominal pixel areas and take into account the detector-based differences in pixel responsivity.

Each observation was thus processed through the JWST calibration pipeline\footnote{Including, crucially, the straylight correction step which both corrects for the MRS cross-artifact \citep[see Figure 4 of][]{argyriou23} and the residual pedestal dark current \citep[see \S 2.5.1 of][]{morrison23}.} up to and including application of the fringe flat (\S \ref{fringeflat.sec}), and then flatfielded using our first-pass pixel flatfield.  No photometric calibration was applied prior to using the JWST 3D drizzle algorithm \citep{law23} to combine the dithered observations into a composite data cube in units of flatfielded DN/s.  We then computed the ratio between the individual flatfielded pixel values and the composite data cube interpolated to the spatial and spectral coordinates of those pixels.  We median combined these values across all 36 exposures (18 per Cycle\footnote{We combined the observations from Cycle 1 and Cycle 2 as there is no clear evidence for evolution between the two.}) to obtain a set of correction factors that we then multiplied by the first-pass flatfield to determine our second-pass flatfield.

This process was then iterated again, using the second-pass flatfield to compute a new composite data cube and derive a second set of corrections to produce the third-pass flatfield.  During this step we compute the root-mean-squared (RMS) scatter between the interpolated cube spaxels at each wavelength and the immediately adjacent four spaxels; if this RMS is more than 20\% of the derived value we ignore this data point when averaging across the different observations.  This excludes the most uncertain values where the on-sky intensity distribution is changing rapidly (e.g., in Channel 4 where the field of view extends beyond the edge of the nebula for some mosaic tiles).
This iterative approach converges rapidly, with corrections in the second-pass flatfield of about 2\% on average, and corrections in the third-pass flatfield of about 0.5\%.

\begin{figure*}[!]
\epsscale{1.1}
\plotone{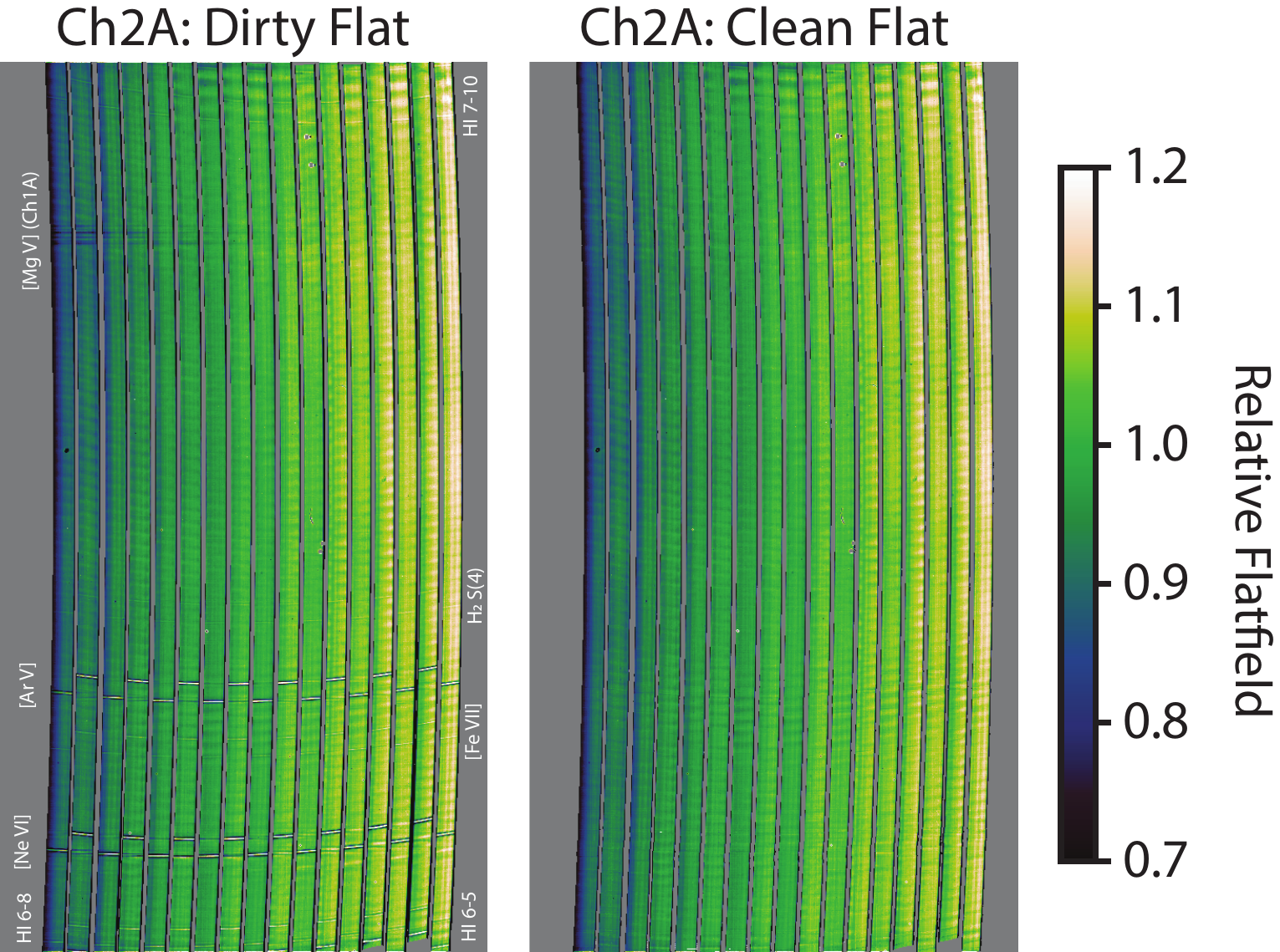}
\caption{Pixel flatfield for MIRI MRS Channel 2A ($\lambda\lambda$ 7.51 -- 8.77 $\micron$) derived from observations of the planetary nebula NGC 7027.  Note the narrow artifacts in the ``dirty'' flat (left panel) due to bright emission lines within the nebula that are patched to create the final ``clean'' flat (right panel).  Prominant spectral artifacts are labelled; dark artifacts near the top of the detector are due to extremely bright [Mg V] emission in Channel 1A on the other half of the detector.  Large-scale variations in flatfield values between and across slices are due to differences in the effective pixel area, while bright/dark periodic spectral banding in the vertical/dispersion direction are due to the pixel flat fixing some residual fringing signal not fully corrected by the fringe flat.
}
\label{pixflat.fig}
\end{figure*}

While this method works well at continuum-dominated wavelengths where the nebula varies smoothly both spatially and spectrally, it is unreliable in the vicinity of emission lines for which undersampling can severely bias the recovered surface brightness comparisons.\footnote{Since NGC 7027 is extremely highly ionized, these lines include both HI transitions (e.g., HI 6-5) as well as species such as [Ar V] and [Ne IV]; see discussion by \citet{bs01}.}  We therefore patch such spectral regions (and others affected by a variety of detector artifacts) in these ``dirty'' flats by hand, interpolating from nearby spectral regions to fill in the missing information.  This interpolation respects visible trends with spatial position along a given slice and known odd/even row and column variations.
Additionally, we trim unreliable edge pixels within each slice, using only those pixels with 50\% or greater relative responsivity to ensure that low-throughput pixels do not contaminate the final data cubes.
Figure \ref{pixflat.fig} shows the final ``clean'' pixel flat for Channel 2A and illustrates the typical features of the flatfield and the necessity for interpolating over spectral emission features in the nebula.

Using the RMS difference between the flatfield corrections provided by the 36 individual exposures, it is possible to 
estimate the uncertainty in the final flatfield for each pixel.
This uncertainty is typically about 1.5\% in Channels 1--3, and about 4\% in Channel 4.  Empirically, we have found that these uncertainties are fairly accurate via an analysis of the noise properties of calibrated MRS standard star spectra (\S \ref{discussion.sec}), indicating that flatfield uncertainties can be a limiting factor for the effective S/N of very bright targets.  At present it is unclear why the flatfields have this limiting uncertainty and whether it is due to our choice of calibration target/method or some innate property of the MRS detectors (e.g., the changing fringe signal between different pointings).

Unfortunately, it was not possible to derive new pixel flatfields for Channels 1A and 1B due to the limited S/N provided by NGC 7027 in these wavelength bands.  We therefore continue to use the ground-based pixel flats described by \citet{argyriou23} in these two bands.\footnote{\citet{argyriou23} combine the flatfield and photometric response vectors into a single 2D reference file; we divide out a 1D spline fit to their wavelength-dependent photometric response to leave just the pixel flatfields.}
This represents a tradeoff in flatfield quality; while the ground-based flats had extremely high statistical accuracy given the brightness of the extended blackbody calibration source, they may have systematic biases resulting from the different cone angle and large-scale illumination patterns.  Likewise, the substantially different fringe patterns observed between on-sky and internal calibration lamp observations limits the utility of such data for improving our understanding of the flatfield appropriate for typical science observations.

\subsection{$11.6 \, \micron$ Artifact}

\begin{figure}[!]
\epsscale{1.1}
\plotone{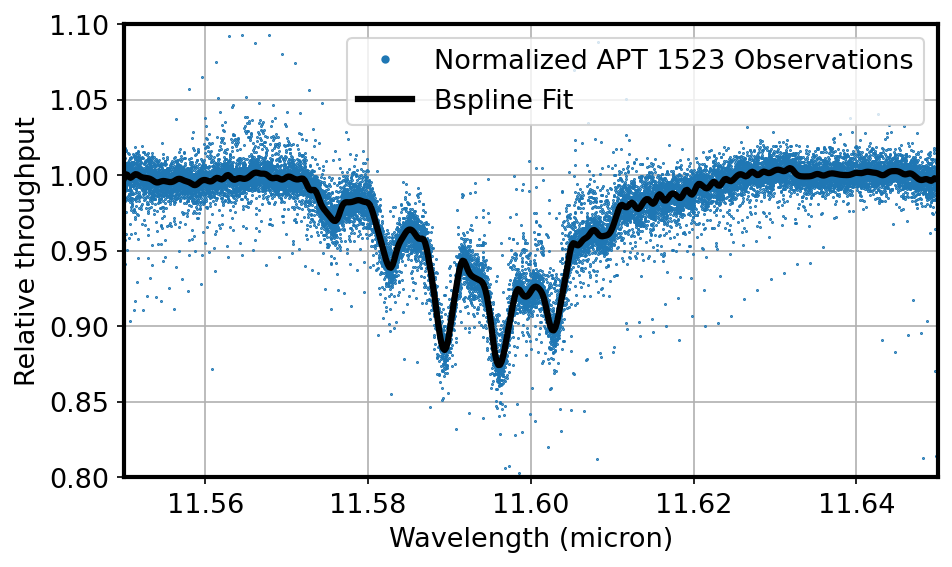}
\caption{Channel 2C spectrum of NGC 7027 (blue points) normalized by a continuum fit that interpolates over the 11.55--11.65 $\mu$m wavelength range.  All detector pixels are plotted individually according to their native wavelength solution, resulting in a dramatic oversampling of the spectrum compared to the nominal spectral resolution of the instrument.  The resulting spectrum (nicknamed the ``bearclaw'') shows strong evidence for systematic absorption due to GeO.  The solid black line represents the cubic b-spline fit incorporated into the pixel flatfield.
}
\label{bearclaw.fig}
\end{figure}

The Channel 2C pixel flats include an additional correction for 
a throughput artifact in the 11.55--11.65 $\micron$ wavelength range (shown in Figure \ref{bearclaw.fig}).\footnote{While initially discovered in flight, this feature has been
found to be consistent with ground test data as well.}
This artifact has been unambiguously identified as due to germanium oxide (GeO) absorption in blocking filter BF2 \citep{wells15}, which is the only element in the Channel 2 optical path that includes a significant path length of germanium.  This feature is not seen in Channel 3A data covering a similar wavelength range, since a blocking filter is not used in Channel 3.

While this feature might more obviously be incorporated into our 
photometric calibration vector, its high-frequency nature lends itself to better correction in the 2D pixel flatfield, allowing the photometric calibration vector to vary smoothly with wavelength.
We fitted the artifact with a high-frequency cubic basis spline \citep[``b-spline'',][]{deboor77}\footnote{We use the ``PyDL'' package \citep{weaver24} implementing spline fitting functions used extensively by the Sloan Digital Sky Survey \citep[e.g.,][]{bolton07,law16}.}, evaluated at the wavelengths of each detector pixel in order to derive the necessary correction.

\section{Time-Dependent Count Rate}
\label{tputloss.sec}

During the first year of observations, we discovered that the MIRI MRS \citep[and to a lesser degree the MIRI imager;][]{gordon24} has a significant time-dependent loss in the detector count rates produced by a fixed input signal.
This effective throughput loss was not observed during ground testing 
or accounted for in the initial calibration vectors derived during commissioning \citep{argyriou23},
and has been a significant complicating factor in deriving a reliable flux calibration for the MRS.
We characterize this loss by comparing regular monthly observations of the MRS internal calibration lamps in all twelve spectral bands (JWST PIDs 1518, 4484, and 6612 for Cycles 1, 2, and 3 respectively) to identical observations obtained during commissioning on 2022 April 11 (JWST PID 1012).

\begin{figure*}[!]
\epsscale{1.1}
\plotone{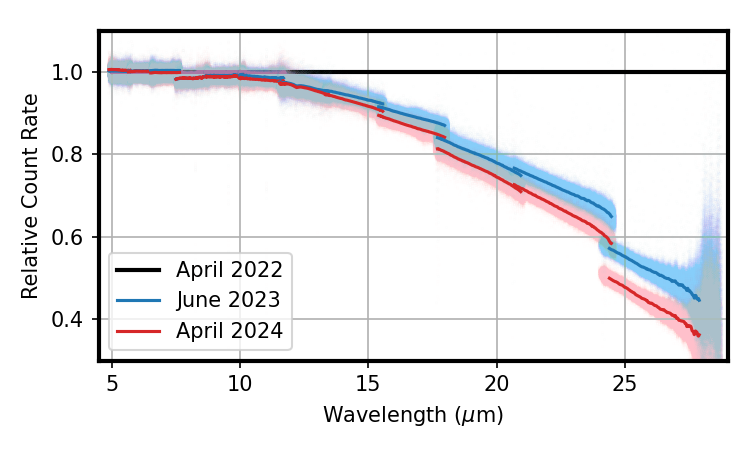}
\caption{Ratio of the count rate produced by identical observations of the MRS internal calibration lamps in June 2023 (blue points) and April 2024 (red points) to observations obtained in April 2022.  Light blue/red points indicate individual detector pixels, while the dark blue/red lines represent b-spline fits to the blue points at the spectral resolution of the MRS.  The discontinuities at overlapping wavelength ranges represent jumps between individual spectral bands.
}
\label{loss_spec.fig}
\end{figure*}

In Figure \ref{loss_spec.fig} we plot the ratio of the calibration lamp spectra
(in units of DN/s) observed in June 2023 and April 2024 to the original April 2022 observations.\footnote{These spectra have been additionally processed to remove residual fringing artifacts and any spatial variations due to slitlet mask contaminants, and isolate only the wavelength-dependent loss at high S/N.}
The throughput loss is most pronounced at longer wavelengths, with  observations at $\lambda = 25 \, \micron$ producing roughly 50\% the count rate on the detector in April 2024 as in April 2022.  The loss spectrum is generally smooth with no evidence for spectral features resolvable by the MRS ($R \sim 3000$) that might be indicative of water ice contamination.  There are, however, discrete jumps at fixed wavelength between adjacent spectral bands.  At $\lambda = 24.5 \, \micron$, for instance, Channel 4B had experienced about 40\% loss of throughput relative to commissioning while Channel 4C had experienced about 50\% loss.  The root cause of this decrease in effective throughput is most likely space weathering caused by post-launch high-energy cosmic ray impacts.  These differences between bands then at least partially implicate the sub-band specific optics in the DGA wheels, while the contribution of other components (including the detectors) are also under investigation.

\begin{figure*}[!]
\epsscale{1.1}
\plotone{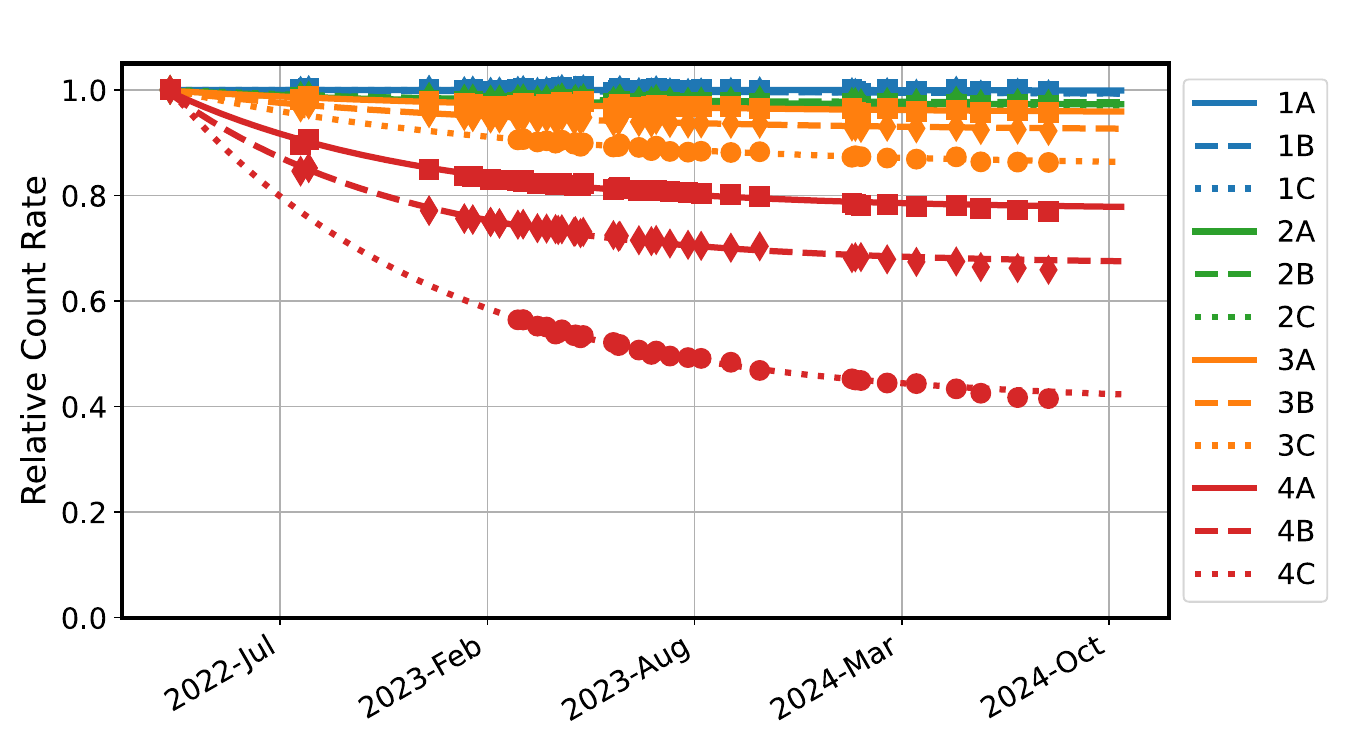}
\caption{Relative count rate produced over time from identical observations of the MRS internal calibration lamps at the median wavelength of each of the twelve MRS spectral bands 1A-4C.  Solid points show measured values while solid/dotted/dashed lines indicate the empirical
model fits.  Large gaps in time coverage appear when the MRS was offline in Fall 2022 for investigation of a (likely unrelated) grating wheel anomaly, and in Winter 2023 when multiple observations
failed for unrelated reasons.  Band 'C' monitoring observations are missing prior to March 2023 due to an error in the PID 1518 program configuration, which complicates the fitting of a loss model without the corresponding Fall 2022 points.
}
\label{timedep.fig}
\end{figure*}

This decrease in effective throughput is not only smooth with wavelength within each spectral band, but with time as well.  Figure \ref{timedep.fig} shows the relative count rate at the median wavelength within each spectral band as a function of time since the original commissioning observation.  As in Figure \ref{loss_spec.fig}, the change is most pronounced at long wavelengths, but at all wavelengths we observe an asymptotic levelling off of the throughput over time.  While Channel 4B lost 18\% effective throughput in the first 6 months, it has lost just 1\% in the 6 months prior to April 2024.
We therefore model the time-dependent instrument throughput $T$ using an empirically-motivated function of the form
\begin{equation}
T = T_0 \, e^{(-b \, [t-t_0]/\tau)} + c
\end{equation}
where $t$ is the modified Julian date (MJD) of observation, $t_0 = 59680$ is the MJD corresponding to April 11, 2022, and $\tau = 100$ days is a characteristic time constant.  $T_0$, $b$, and $c$ are all unknown constants to be constrained by observations.
We calculate such models first for all data within each of the 12 wavelength bands to bound reasonable ranges for parameters $T_0$, $b$, and $c$, and then repeat this fit for ten equally-spaced wavelength bins within each of the bands in which the parameters are allowed to vary
only slightly about the initial solution.

By interpolating these models to the MJD of a given observation, and the wavelength of each detector pixel within each spectral band, we can therefore correct all MRS data back to the zero-day count-rate that they would have produced on April 11, 2022.  The JWST calibration pipeline is designed to automatically apply this correction to all observations as part of the photometric calibration step.

In practice, our understanding of the count rate loss continues to evolve, and periodic updates are made to the model coefficients as additional observations allow us to better constrain the loss function.


\section{Aperture Correction}
\label{apcorr.sec}

The field of view of the MRS is relatively small compared to the size of the point-spread function (PSF), particularly in Channel 4 (field size 7 arcsec vs PSF FWHM 1 arcsec at 25 $\micron$).  At the same time, a relatively high thermal background necessitates the use of annular background subtraction within this field to obtain reliable background-subtracted fluxes for observations of point sources.  Correction factors for the finite size of the extraction aperture and the fraction of the PSF removed via the annular background subtraction\footnote{Typically about 0.5\% for Channels 1--2, 1\% for Channel 3, and 2\% for Channel 4; see Figure 15 of \citet{argyriou23}.} are thus essential for proper flux calibration of the resulting data.

\begin{figure}[!]
\epsscale{1.2}
\plotone{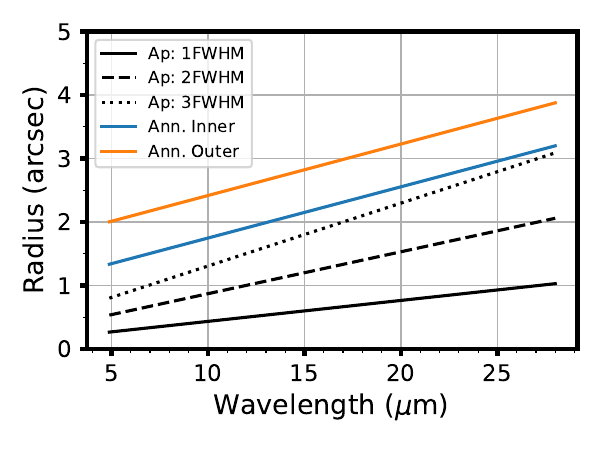}
\caption{Default extraction aperture radius (blue line), and inner/outer background annulus radii (orange/green lines respectively) for MRS point-source spectral extraction.
}
\label{aperture.fig}
\end{figure}

The JWST pipeline applies traditional imaging aperture photometry for point source observations, performing circular aperture extraction and annular background subtraction at each wavelength plane of the 3D data cubes.  Both the aperture extraction radius and inner/outer radii of the background annulus grow linearly with wavelength, resulting in a tapered conical aperture as the effective MRS PSF changes by roughly a factor of five due to diffraction over the 5--25 \micron\ wavelength range (Figure \ref{aperture.fig}).  The default radius is chosen to be  twice the PSF FWHM to minimize resampling artifacts \citep[see discussion by][]{law23}, while the annular background region is chosen to be as large as possible while still fitting within the MRS field of view for a single pointing.

\begin{figure}[!]
\epsscale{1.2}
\plotone{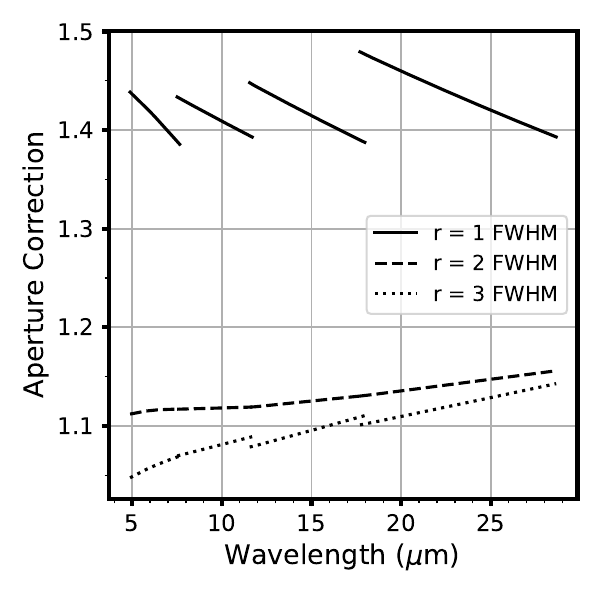}
\caption{MRS aperture-correction factors for point-source aperture photometry with extraction radii of 1, 2, and 3 times the PSF FWHM.  Approximately 5--10\% of the total flux is lost outside the IFU field of view.
}
\label{apcorr.fig}
\end{figure}

We used high-resolution models of the MRS PSF (P. Patapis et al. 2024 in prep.) to derive the necessary correction factors at the default extraction radius $r = 2$ FWHM, and find values of about 10--15\% (dashed line in Figure \ref{apcorr.fig}).\footnote{The aperture correction factors shown here correspond to the jwst\_miri\_apcorr\_004 -- 007 reference files available on CRDS (https://jwst-crds.stsci.edu/).}  We also estimated the relevant aperture correction factors for non-standard aperture radii ranging from $r = 0.5 - 3.0$ FWHM using a combination of high-resolution models and on-sky observations of bright point sources.  The JWST calibration pipeline automatically interpolates between this grid of correction factors for any arbitrary radius within this range.  As illustrated in Figure \ref{apcorr.fig}, some of these correction factors are discontinuous between spectral channels since the differences in spatial sampling have an appreciable impact on the structure of the reconstructed PSF in the data cubes.

\section{Photometric Calibration: Channels 1-3}
\label{photom123.sec}

With flatfields, the time-dependent count-rate loss, and aperture correction factors taken care of, we derived the 1-D photometric calibration vector necessary to convert instrumental count rates in flatfielded DN/s to physical units of MJy/sr for each spectral band.
We discuss first the case of Channels 1--3 ($\lambda < 18$ \micron), for which current JWST spectrophotometric calibration stars sufficed to obtain a reliable calibration.

\subsection{Photometric Calibration}


Table \ref{results.table} lists the ten different calibration stars selected from \citet{gordon22} that have been observed in all of the MRS wave bands.
Nine of these stars were observed as part of the formal
cross-instrument flux calibration program (Cycles 1 and 2 program IDs 1536, 1538, 4496, 4497, 4498) and used a standard four-point dither pattern optimized for point sources at all wavelengths.  One additional target was obtained as part of the MRS geometric distortion and PSF calibration program (program ID 1524), which dithered the bright O9 V star 10 Lac using a customized 57-point dither pattern designed to characterize the distortion solution throughout the field of view \citep{patapis24}.

We processed each set of observations through the standard {\sc calwebb\_detector1} pipeline to produce rate images in units of DN/s for each exposure, and through the {\sc calwebb\_spec2} pipeline up to and including the straylight, static fringe, and pixel flatfield calibration steps.  As in \S \ref{pixflat.sec} the straylight step includes subtraction of a kernel-convolved model for the detector scattered light (i.e. the `cross artifact'), along with subtraction of the residual pedestal dark current determined from the median count rate in the region between the two MRS channels on each detector that sees no direct light from the sky.  This latter correction is essential as the pedestal can be $\sim$ 0.3 DN/s, which is comparable to the signal from the thermal background in Channel 4 ($\sim$ 0.4 -- 1.0 DN/s).

After these standard steps, we applied the time-dependent count-rate correction derived in \S \ref{tputloss.sec} to correct all observations back to the zero-day count rate that they would have produced on April 11, 2022.  After this step we combined the resulting data into data cubes for each of the twelve individual bands using the 3-D drizzle algorithm \citep{law23}, and extracted 1-D spectra from these cubes using the standard 1-D aperture extraction with annular background subtraction  and aperture correction at each wavelength plane (see \S \ref{apcorr.sec}).  We use the default conical apertures in which the extraction radius is equal to twice the PSF FWHM (see Figure \ref{aperture.fig}), and apply the optional arguments ``ifu\_autocen'' to ensure that the extraction aperture is precisely centered on the star in each data cube and ``ifu\_rfcorr'' to apply the 1-D residual fringe correction (P. Kavanagh et al., 2024, in prep.).
The results are uncalibrated source spectra, and the ratio between these and the corresponding
stellar spectral model gives a (noisy) spectrophotometric calibration vector.

\begin{figure*}[!]
\epsscale{1.1}
\plotone{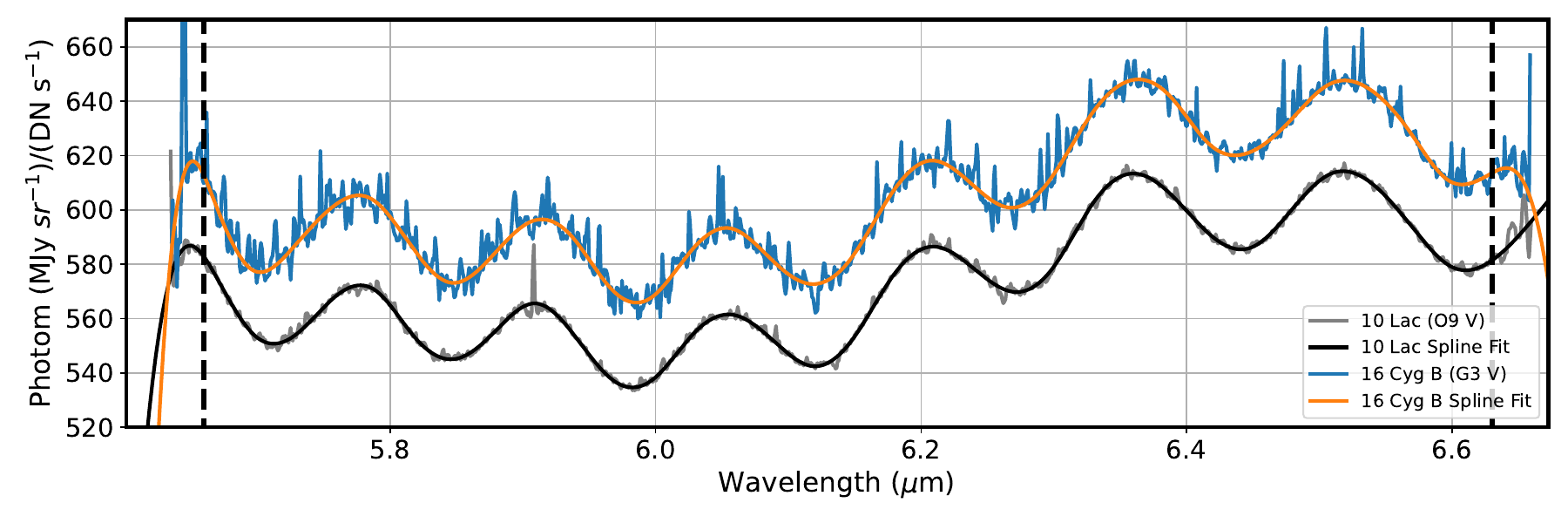}
\caption{Photometric calibration vectors for Channel 1B derived from observations of O9V star 10 Lac, and G3V star 16 CygB.  Grey/blue lines show the ratio of the model to observed stellar spectrum for 10 Lac/16 CygB respectively, while the black/orange lines show the basis spline fits.  The dashed vertical lines indicate the wavelength limits used by the standard pipeline to build spectral cubes for this band; outside these limits partial spatial coverage of the field makes the spectra unreliable.  The vertical offset between calibration vectors derived from 10 Lac vs 16 Cyg B is real.
The emission feature in the 10 Lac spectrum at 5.908 $\micron$ is the HI 9-6 (Humphreys $\gamma$)
transition.
}
\label{photom1b.fig}
\end{figure*}

We adopted the latest STScI CALSPEC spectral models \citep[Table \ref{results.table};][]{bgt14, bohlin22}\footnote{https://www.stsci.edu/hst/instrumentation/reference-data-for-calibration-and-tools/astronomical-catalogs/calspec} for our analysis.
The corresponding ratios between the uncalibrated spectra and the CALSPEC models are shown in Figure \ref{photom1b.fig} for the example case of 10 Lac and 16 Cyg B in Channel 1B.  As illustrated in this figure the direct ratio of uncalibrated spectra and CALSPEC model is noisy, 
and also shows periodic modulations in some bands due to low-frequency fringing caused by the detector
buried contact that are not currently corrected by the static fringe flats \citep{argyriou20b}.\footnote{In the future, this periodic component may instead be incorporated into the fringe flats.}
We therefore fit this ratio with a basis-spline function to determine the best-fitting calibration factor for each star.  We use the minimum number of spline breakpoints necessary to convincingly fit broad variations in the spectral response function while iteratively rejecting and interpolating over narrow spectral features and any residual fringing artifacts.  Likewise, we use these response functions to define the default wavelength cutoffs used for IFU cube building in the JWST pipeline, trimming those wavelengths where the calibration vectors from different stars start to diverge rapidly (dashed vertical lines in Figure \ref{photom1b.fig}).

\begin{deluxetable*}{cccccccccccccccccc}
\label{results.table}
\tablecolumns{12}
\tablewidth{0pc}
\tabletypesize{\small}
\tablecaption{Photometric Calibration Observations}
\tablehead{
\colhead{} & \colhead{} & \colhead{} & \colhead{} & \colhead{} & \colhead{CALSPEC} & \multicolumn{12}{c}{S/N\tablenotemark{a}} \\
\colhead{Source} & \colhead{Type} & \colhead{PID} & \colhead{Observation} & \colhead{Date} & \colhead{Version} & \colhead{1A} & \colhead{1B} & \colhead{1C} & \colhead{2A} & \colhead{2B} & \colhead{2C} & \colhead{3A} & \colhead{3B} & \colhead{3C} & \colhead{4A} & \colhead{4B} & \colhead{4C}
}
\startdata
10 Lac\tablenotemark{b} & O9 V & 1524 & 16 & Jul 2022  & 5 & 975 & 1190 & 302 & 656 & 621 & 703 & 894 & 995 & 752 & 476 & 208 & 35\\
$\mu$ Col & O9.5 V & 4497 & 4 & Dec 2023 & 5 & 543 & 654 & 249 & 346 & 404 & 402 & 342 & 351 & 315 & 118 & 52 & 10\\
$\delta$ UMi & A1 V & 1536 & 24  & Apr 2023  & 5 & 453 & 536 & 374 & 376 & 444 & 439 & 584 & 716 & 659 & 445 & 214 & 45\\
HR 5467 & A1 V & 4496 & 9 & Mar 2024 & 4 & 534 & 639 & 431 & 335 & 422 & 379 & 304 & 364 & 341 & 128 & 54 & 9\\
HD 2811 & A3 V & 1536 & 22  & Nov 2022  & 5                    & 501 & 539 & 344 & 340 & 393 & 337 & 399 & 362 & 257 & 157 & 51 & 10\\
HD 163466 & A7 Vm & 1536 & 23  & Mar 2023   & 4                & 635 & 722 & 373 & 365 & 502 & 392 & 399 & 411 & 316 & 214 & 73 & 13\\
HR 6538 & G1 V & 4498 & 11 & Jun 2024 & 7 & 98\tablenotemark{c} & 265 & 365 & 365 & 470 & 383 & 299 & 389 & 347 & 119 & 59 & 11\\
16 Cyg B & G3 V & 1538 & 1  & Aug 2022   & 5                   & 84\tablenotemark{c} & 244 & 397 & 390 & 508 & 422 & 421 & 504 & 542 & 202 & 182 & 42\\
HD 37962 & G3 V & 1538 & 2  & Mar 2023 & 9                    & 99\tablenotemark{c} & 364 & 431 & 450 & 458 & 377 & 387 & 418 & 339 & 186 & 77 & 13\\
HD 167060 & G3 V & 1538 & 3  & Aug 2022 & 6                   & 96\tablenotemark{c} & 282 & 340 & 374 & 388 & 345 & 356 & 346 & 265 & 164 & 57 & 15\\
515 Athalia & Asteroid & 1549 & 6 & Apr 2023 & ... & 13 & 43 & 127 & 252 & 377 & 485 & 727 & 799 & 737 & 658 & 464 & 165 \\
SAO 206462 & YSO & 1282 & 55 & Aug 2022 & ... & ... & ... & ... & ... & ... & ... & 509 & 659 & 943 & 711 & 713 & 310 \\
\enddata
\tablenotetext{a}{Bootstrapped S/N per wavelength element.}
\tablenotetext{b}{Observed with a 57-point dither pattern to constrain distortion model during Cycle 1.}
\tablenotetext{c}{Bootstrapped estimates are low due to CO band features in the spectrum.}
\end{deluxetable*}

Table \ref{results.table} shows that
the effective S/N\footnote{Estimated from a bootstrapped analysis of the spectra themselves by calculating the RMS variations when the calibrated spectrum is divided by a moderate-order basis spline fit to the spectrum to remove spectral features and any leftover residual fringing.} of these ten stars can vary considerably in different bands.  The spectrum of 10 Lac has by far the highest S/N (typically 600--1000 throughout Channels 1A--3C; see also Figure \ref{photom1b.fig}) thanks to its intrinsic brightness, total integration time, and 57-point dither pattern marginalizing any flatfield systematics across the entire field of view.
We therefore derive our initial Channels 1A--3C response function from the 10 Lac calibration vector as the bright O-dwarf spectrum contains only moderate spectral features easily interpolated over by the spline routine.

\begin{figure*}[!]
\epsscale{1.1}
\plotone{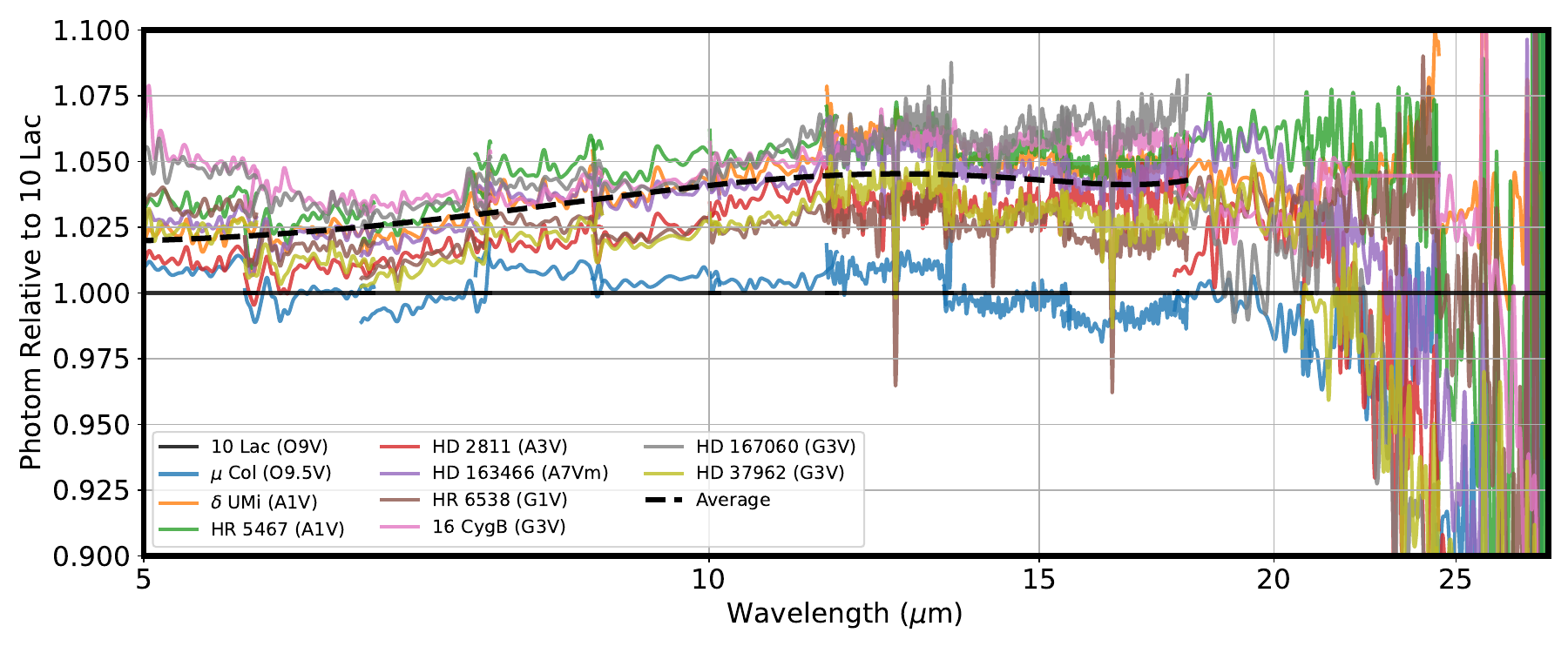}
\caption{Relative photometric calibration vectors of individual stars (colored lines) compared to that found for 10 Lac.  Larger values correspond to stars that are fainter than the model and thus need larger correction factors to bring into agreement; note that discontinuities can occur between spectral bands.  The dashed black line shows a low-order polynomial fit to the average of all ten stars at $\lambda < 18$\micron; we adjust the calibration vector derived from 10 Lac by this model average. 
}
\label{relphotom.fig}
\end{figure*}

Despite the high SNR of the 10 Lac spectrum, model predictions for any individual star typically have systematic uncertainties $\sim$ 2\% \citep{bgt14}.  Indeed, the
calibration vectors derived from each of the different stars (Figure \ref{relphotom.fig}) show clear systematic offsets in addition to higher frequency variations due to differences in the spline model fit to the noisier spectra.
We therefore fit a low-order polynomial to the average response vectors of all ten stars (dashed black line in Figure \ref{relphotom.fig}), and multiply the 10 Lac calibration vector by this polynomial in order to obtain the final calibration vector for each band.


The resulting spectrophotometric calibration vectors are combined with the time-dependent loss models for each band and applied together by the JWST calibration pipeline during the flux calibration step in the {\sc calwebb\_spec2} pipeline.
The corresponding flux-calibrated spectrum of 10 Lac is shown in Figure \ref{10lac.fig}, and the continuum-normalized spectrum in Figure \ref{spectrum_10lac.fig} (see Appendix \ref{appendix.sec} for flux-calibrated spectra of the other nine stars in our sample).
As was known from prior Spitzer IRS observations \citep{marcolino17}, the hydrogen emission lines present in 10 Lac are relatively weak and the spectrum is otherwise smooth and regular (with the exception of some broad emission-line features that we discuss in \S \ref{nev.sec}) with no obvious red excess longward of 18~$\micron$ indicating an
unresolved dust disk.

\begin{figure*}[!]
\epsscale{1.1}
\plotone{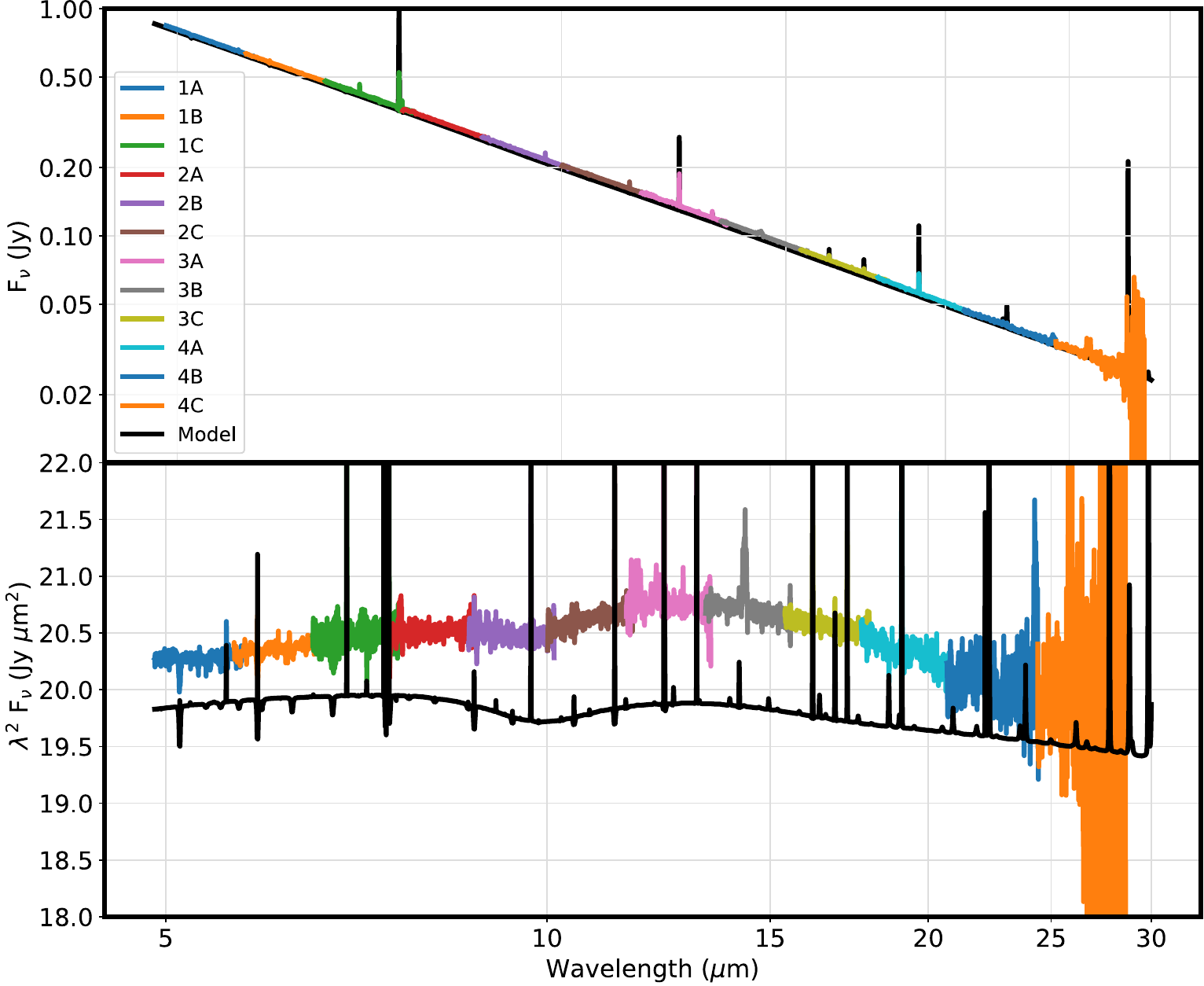}
\caption{Flux-calibrated spectrum of O9 V star 10 Lac.  Individual colored segments represent spectra from  specific MRS bands while the solid black line represents the CALSPEC model.  Note that the 10 Lac spectrum deviates systematically from the model as the MRS photometric calibration is normalized to the average of all standard stars.
}
\label{10lac.fig}
\end{figure*}

\begin{figure*}[!]
\epsscale{1.1}
\plotone{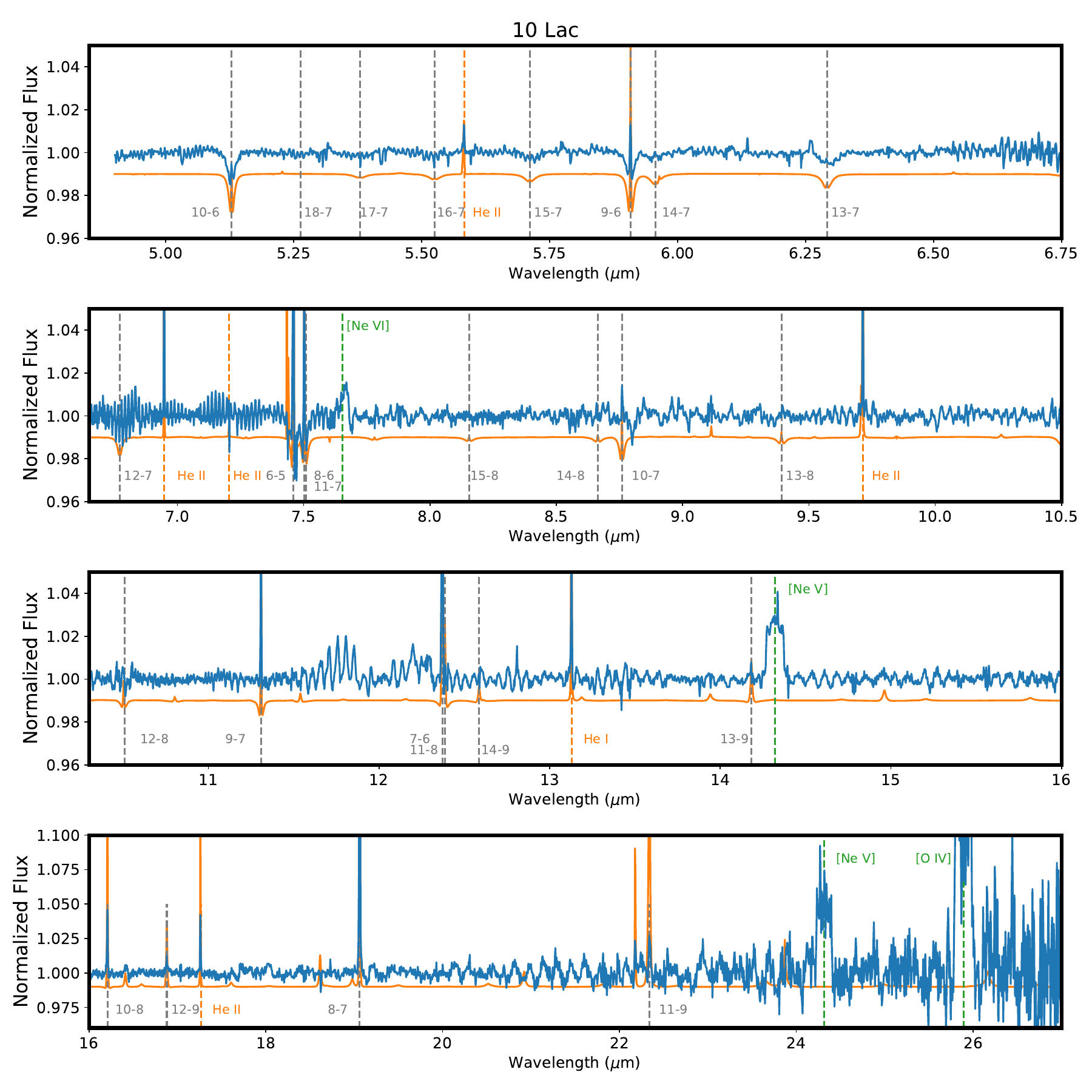}
\caption{Continuum-normalized spectrum of 10 Lac (solid blue line) and the corresponding CALSPEC model (solid orange line, vertically offset for clarity).  Note that many wavelengths show sinusoidal artifacts at the $\sim 0.5$\% level due to fringing that has not been completely corrected by the JWST calibration pipeline.  Dashed grey lines indicate major H transition lines (labeled by quantum level), while dashed orange lines indicate He transitions.  Note that the bottom panel has a larger vertical scale.
}
\label{spectrum_10lac.fig}
\end{figure*}

\subsection{Broad High-Ionization Emission Lines}
\label{nev.sec}

\begin{figure}[!]
\epsscale{1.1}
\plotone{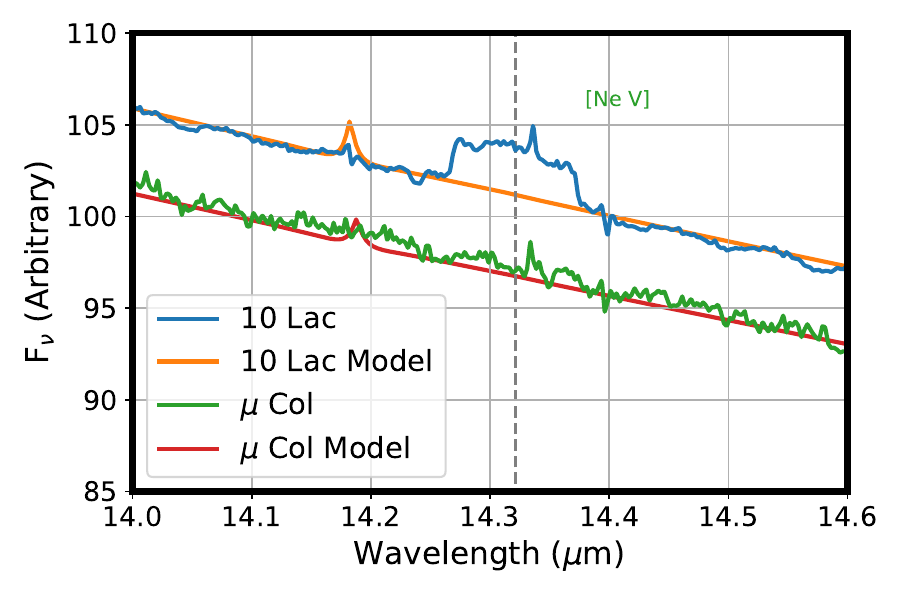}
\caption{Calibrated spectra and CALSPEC models for $10$ Lac (O9 V; blue/orange lines respectively) and $\mu$ Col (O9.5 V; green/red lines respectively, scaled by a factor of 1.5) illustrating a broad emission feature in the 10 Lac spectrum that we identify as due to \nev\ $\lambda = 14.3217 \micron$.  This feature is not observed in $\mu$ Col, which has a similar spectral class.
}
\label{14mbox.fig}
\end{figure}

While the 10 Lac CALSPEC model is generally an excellent match to the observed spectrum, one notable exception is an unusual broad, flat-topped emission feature 
between 14.27 and 14.37 $\micron$ (Channel 3B).  We tentatively identify this feature as due to
\nev\ $\lambda = 14.3217 \micron$ emission, with corresponding \nevi\ $\lambda 7.6524 \micron$, \nev\ $\lambda 24.3175 \micron$, and \ofour\ $\lambda 25.89 \micron$ as well.  These features are discussed further by \citet{law24}; for our present
calibration purposes we simply patch the calibration vector in these wavelength ranges with an appropriately
scaled version of the calibration factor derived from O9.5 V star $\mu$ Col instead which does not exhibit
this feature (see, e.g., Figure \ref{14mbox.fig}).

\subsection{12 $\micron$ Spectral Leak}
\label{leak.sec}

Calibration of Channel 3A also required special handling during the calibration process and in the JWST pipeline due to a small spectral leak in the MRS dichroic chain known from ground testing.  Ordinarily, the combination of dichroics and blocking filters on the two DGA wheels ensures that only first-order dispersed light is recorded on the MRS detectors.  As discussed by \citet{argyriou23} however, roughly 2.5\% of the light from 6.1$\micron$ (Channel 1B) is present in second-order at 12.2$\micron$ (Channel 3A), contaminating the observed spectra.

\begin{figure}[!]
\epsscale{1.1}
\plotone{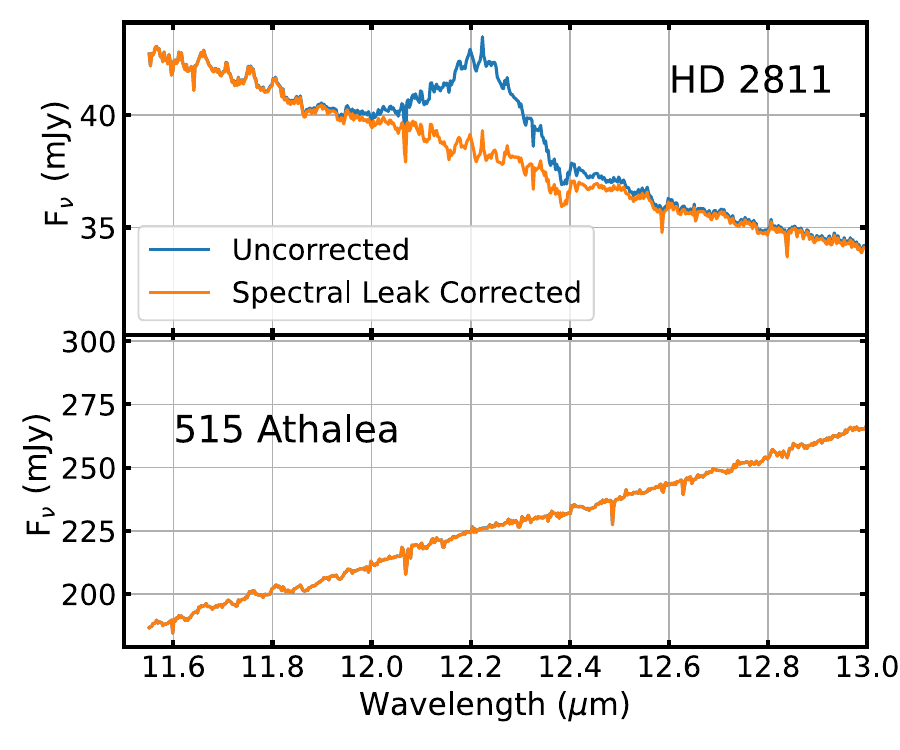}
\caption{1-D spectra of the A3 V standard star HD2811 (top panel) and main-belt asteroid 515 Athalia (bottom panel) illustrating the impact of the 12 $\micron$ spectral leak, which manifests as a fraction of the signal at 6.1$\micron$ being added to the 12.2 $\micron$ flux.  An automated correction for this leak is available in the JWST calibration pipeline; while no correction is necessary for red sources such as 515 Athalia with minimal 6.1 $\micron$ flux, correction is essential for blue sources such as HD 2811.
}
\label{spectral_leak.fig}
\end{figure}

We illustrate this effect in Figure \ref{spectral_leak.fig}; while it is negligible for red sources with little 6 $\micron$ signal, it is pronounced for blue sources and can result in an anomalous broad emission feature around 12.2$ \micron$.  We have calibrated the response function of this second-order leak, and removed it from the observational data prior to deriving the photometric calibration vectors described here.  This response function has been incorporated into the JWST calibration pipeline\footnote{Note that early versions of the pipeline and calibration reference files during the first year of the mission did not account for this artifact, and could correspondingly produce either positive or negative artifacts in the resulting spectra depending on the shape of the input spectrum.} which includes a step to estimate the strength of this signal in Channel 3A based on Channel 1B observations (if present) and subtract it from the corresponding 1-D extracted spectra.  No such general correction is available for the 3-D data cubes or spectra of extended sources however, since Channel 1B observations have a smaller field of view than Channel 3A.

\section{Photometric Calibration: Channel 4}
\label{photom4.sec}

While current JWST standard stars sufficed to flux calibrate Channels 1--3, they are inadequate for reliable calibration of Channel 4 at $\lambda > 18$ \micron.  At such wavelengths the time-dependent throughput loss is large and changing most rapidly (see \S \ref{tputloss.sec}), the telescope thermal background is rising rapidly, and even 5th-magnitude standard stars become extremely faint.  
The stellar-derived photometric response functions illustrated in Figure \ref{relphotom.fig} 
longward of 18 $\micron$ thus show significant variations (likely due to uncertainties in the background subtraction) and create correspondingly large unphysical artifacts in the spectra of bright red sources.
The ongoing JWST calibration program is therefore investigating the suitability of brighter stars for calibration of Channel 4 in observing Cycle 4 and beyond.  At present, however, 
we instead derive the Channel 4 flux calibration from observations of asteroid 515 Athalia 
\citep[PID 1549 Observation 6; observed as a calibrator for the JDISCS program by][]{pontop24}
and the young stellar debris disk SAO 206462 (PID 1282 Observation 55; PI T.\ Henning).

\begin{figure*}[!]
\epsscale{1.1}
\plotone{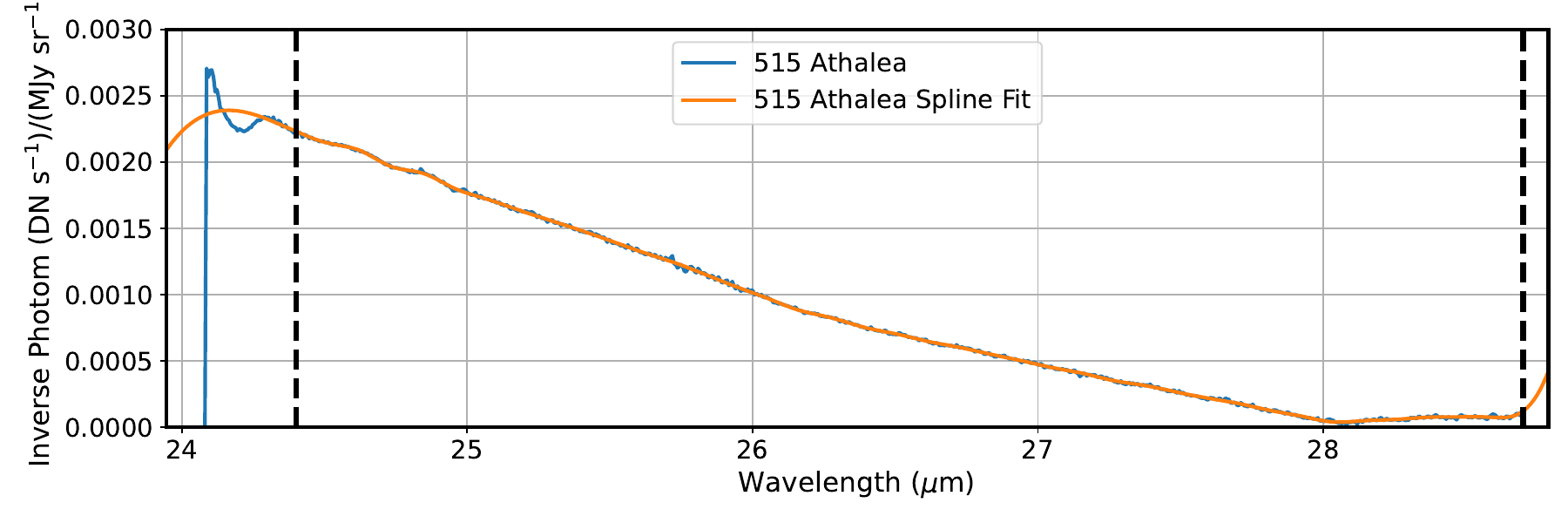}
\caption{As Figure \ref{photom1b.fig}, but for Channel 4C using observations of main-belt
asteroid 515 Athalia.  In this case, however, the spline fit was performed in inverse-photom units to reduce
artifacts that can occur when the effective response dips to near zero.
}
\label{photom4c.fig}
\end{figure*}

\begin{figure*}[!]
\epsscale{1.1}
\plotone{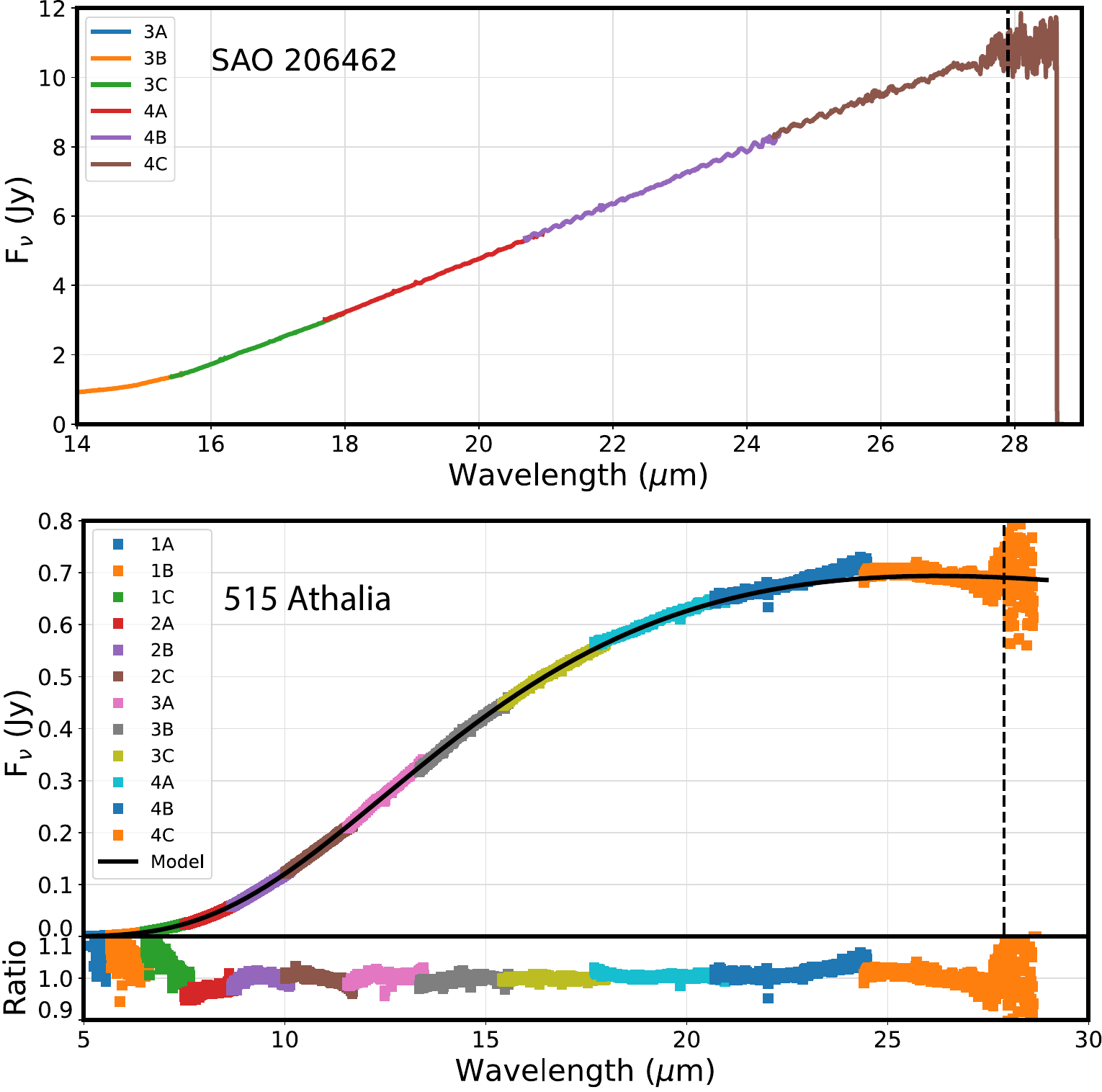}
\caption{Top panel: Calibrated spectrum of the circumstellar disk SAO 206462 in Channels 3 and 4.  Bottom panel: Arbitrarily renormalized spectrum of the asteroid 515 Athalia (see text), overplotted with a best-fit $T= 193.9$ K blackbody model.  In both plots the vertical dashed black line indicates the nominal Ch 4C wavelength cutoff of 27.9 $\micron$, above which calibration quality decreases rapidly.  The lower inset shows the ratio between the renormalized 515 Athalia spectrum and the best-fit model.
}
\label{ch4photom.fig}
\end{figure*}

Starting with the flux-calibrations derived in Section \ref{photom123.sec}, we produce a calibrated spectrum of main-belt Themistian asteroid 515 Athalia.  Since 515 Athalia is itself variable with a 10.6 hour period \citep{pilcher15}, the individual band 1A-3C spectral segments are offset from each other by $\sim$10\%.  Assuming that the temperature of the asteroid is roughly constant and that the variations are due to differences in the effective albedo as it tumbles, we derive a series of empirical scaling factors necessary to align the spectra of bands 1A--3C.  We fit a smooth blackbody curve to the resulting spectrum, finding a best-fit blackbody temperature of 193.9 K (Figure \ref{ch4photom.fig}, lower panel).  This gives us a nominal spectral model of 515 Athalia throughout the 18--28 $\micron$ range; combining this with the observed uncalibrated asteroid spectra we derive the relative photometric response function of bands 4A--4C (see Figure \ref{photom4c.fig}).

Since 515 Athalia is a variable source, it can be used to calibrate the relative response within a given band but not the absolute response compared to the adjacent spectral bands.  We therefore use the extremely bright red debris disk SAO 206462 to bootstrap the appropriate band-to-band normalization using the spectral overlap between adjacent bands.  Channel 4A, for instance, overlaps with Channel 3C in the wavelength range 17.7 -- 17.95 $\micron$; we scale the photometric response of Channel 4A such that the spectrum of SAO 206462 agrees with the Channel 3C calibration.  Likewise, having thus calibrated Channel 4A we use the 20.7--20.95 $\micron$ overlap region to normalize the calibration for Channel 4B, and the 24.4--24.5 $\micron$ overlap to normalize the calibration for Channel 4C.

During this process, we also apply a low-order polynomial correction to the 515 Athalia-based photometric correction in Channel 4B to remove an abnormal downturn longward of 23 $\micron$ that may be due to background subtraction systematics and is not seen in other sources.
Likewise, in Channel 4C
at wavelengths longer than $27.5 \micron$ even the 515 Athalia spectrum becomes unreliably faint as the system throughput falls to near zero.  We therefore assume on the basis of its 18--27 $\micron$ spectrum that SAO 206462 continues to rise smoothly to the Channel 4C wavelength cutoff at 28.8 $\micron$ and adjust the photometric calibration vector accordingly.
Figure \ref{ch4photom.fig} plots the resulting flux-calibrated spectra of SAO 206462 and 515 Athalia and demonstrates that both appear to be well-behaved.

\section{Repeatability and Uncertainty}
\label{repeat.sec}

We assess the repeatability of the MRS photometric calibration using observations of the late A dwarf HD 163466 (A7 Vm).
As noted by \citet{gasman23} and \citet{argyriou23}, this star was one of the first observed during on-orbit commissioning (PID 1050; PI B.\ Vandenbussche) and was used to define the MRS flux calibration for the first year of the mission.
Since that time, HD 163466 has been reobserved regularly as part of a dedicated cross-instrument calibration monitoring program (PIDs 1536, 1539, 4499, 6607).  Initially these monitoring observations focused on band A; with the discovery of the time-dependent throughput loss described in \S \ref{tputloss.sec} the observations were expanded to include bands B and C as well.
In total, there have been 21 unique observations of this source between June 2022 and July 2024.

We process the data from all of these observations through the standard JWST pipeline, using the time-dependent correction models and photometric calibration vectors derived above.  In this processing we create individual per-band spectra for each of the observations, applying the ``ifu\_autocen'' aperture centroiding and the ``ifu\_rfcorr'' 1d residual defringing routine.

\begin{figure*}[!]
\epsscale{1.1}
\plotone{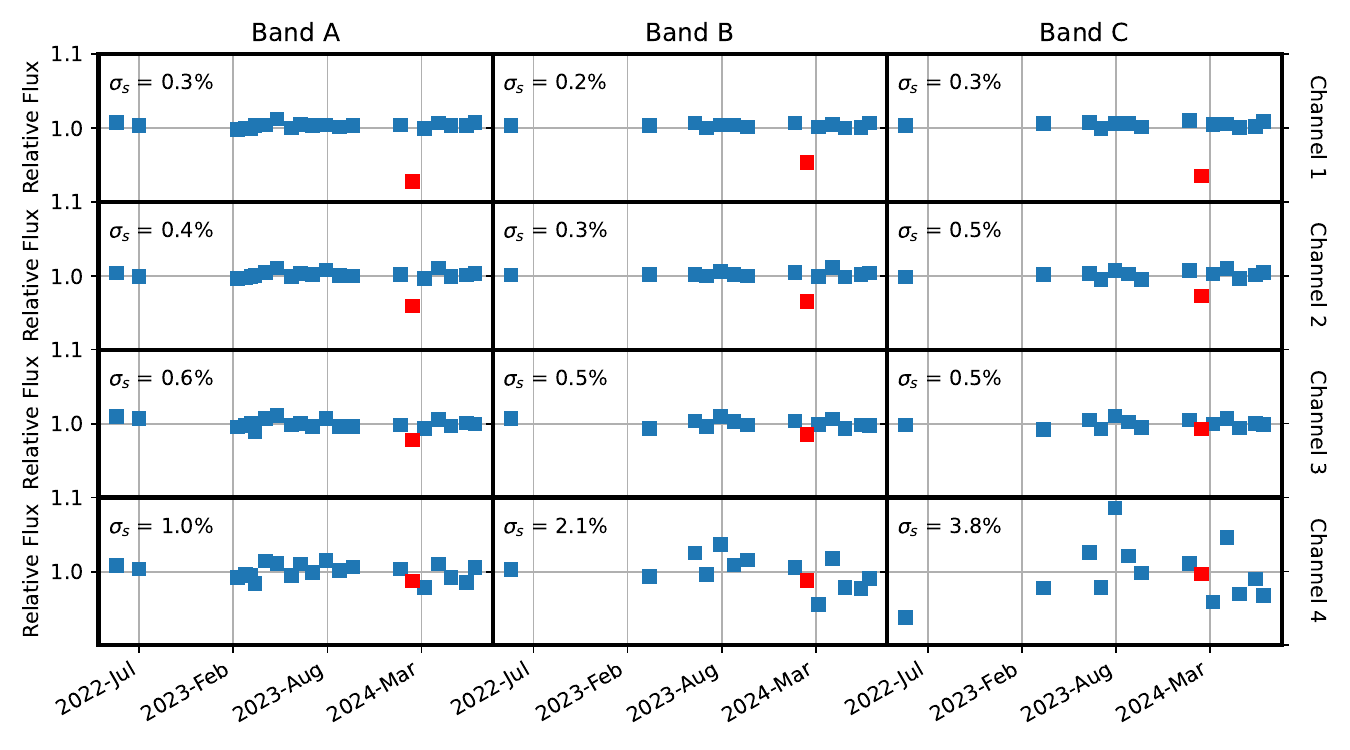}
\caption{Systematic repeatability for all twelve MRS spectral bands based on regular observations of HD 163466 (A7 Vm).  Filled blue squares show the median flux ratio between repeated spectra in each band from June 2022 to July 2024.
Inset text values indicate the repeatability (defined as the RMS scatter $\sigma_s$)  between the individual measurements.
The repeatability is below 1\% in Channels 1A--3C, while the greater scatter in channel 4 is driven by partly by increasing uncertainty in the background subtraction.
The point highlighted in red was observed on February 28 2024 during a major primary mirror tilt event in which the wavefront error degraded to 327 nm; we exclude this point from our statistics.
}
\label{repeat.fig}
\end{figure*}

In each of these bands we average the individual spectra to construct a high-S/N mean spectrum for the source, and compute the median of the ratio between the individual spectra and the mean spectrum.
This allows us to assess any evolution in the overall normalization of the flux calibration over time; we plot the results in Figure \ref{repeat.fig}.  Despite the large evolution in the effective system throughput over the last two years (\S \ref{tputloss.sec}), the pipeline-processed spectra are consistent to better than 1\% in Channels 1--3 and to within a few percent in Channel 4 (likely due to increasing errors in the background subtraction). 
This scatter may represent the limitations of our smooth time-dependent correction models (i.e., if the throughput loss is discrete on small time scales), but the residuals show no evidence of a temporal trend at any wavelength.

\begin{figure*}[!]
\epsscale{1.1}
\plotone{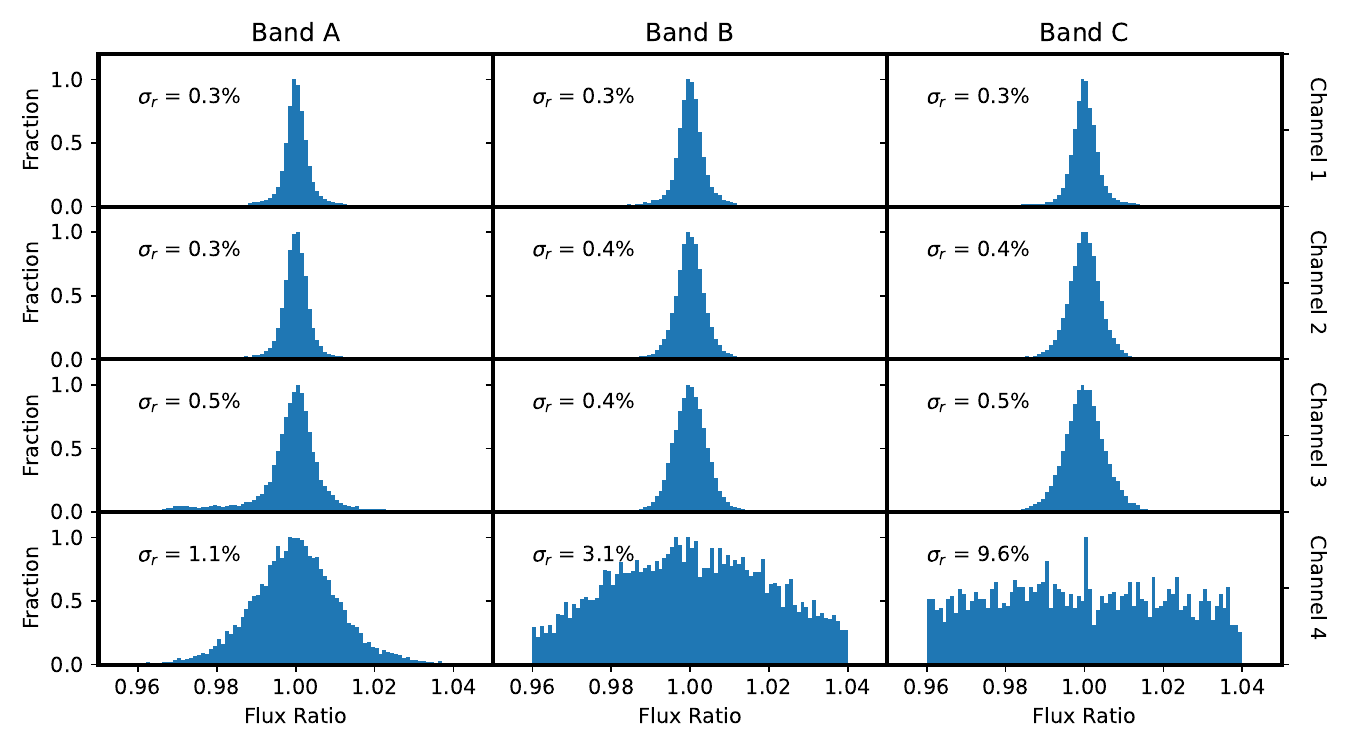}
\caption{Stochastic noise for all twelve MRS spectral bands based on regular observations of HD 163466 (A7 Vm).  Blue histograms represent the range of values about the median for each wavelength element in the extracted spectra, after correction for systematic per-band offsets.  The RMS of these histograms
($\sigma_r$) describes the effective noise in the spectra at a given wavelength
across repeated observations.
}
\label{repeat2.fig}
\end{figure*}

If we correct for the systematic variations in the flux calibration in each band (on the order of a fraction of a percent), the repeated observations can also be used to calulate an empirical S/N ratio from the remaining stochastic noise.
We divide each spectrum by the median spectrum constructed above, normalize by the known systematic flux calibration offset shown in Figure \ref{repeat.fig}, and calculate the resulting distribution of values at each wavelength.
Figure \ref{repeat2.fig} shows these distributions; the characteristic width of the distributions indicates that the ``true'' S/N of any given observation ranges from about 330 per spectral element in Channel 1 to about 10 per spectral elements in Channel 4C.

\begin{figure}[!]
\epsscale{1.3}
\plotone{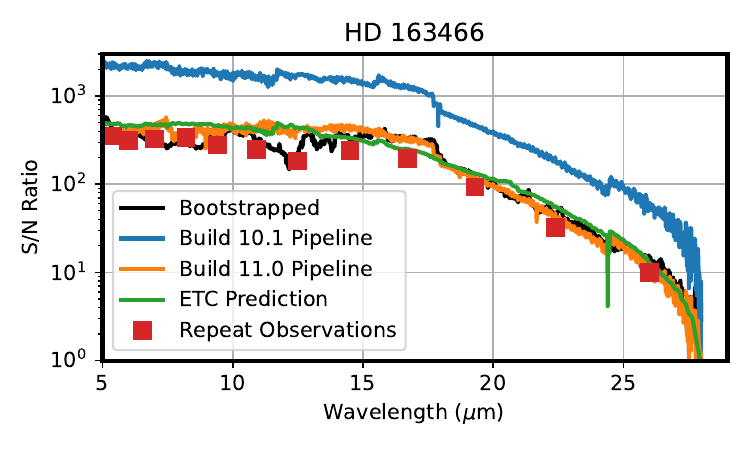}
\caption{S/N ratio of the MIRI MRS spectrum of HD 163466 (A7 Vm) obtained by PID 1536.  The solid blue/orange lines represent estimates provided by the JWST calibration pipeline corresponding to Builds 10.1 and 11.0 respectively, while the expected S/N based on the Cycle 3 Exposure Time Calculator (ETC v3.0) is shown as a green solid line.
The solid black line represents an empirical estimate bootstrapped from rms variations with wavelength in the extracted spectrum, while the red filled squares represent empirical estimates based on repeated observations throughout the first two years of the mission.
}
\label{errplot.fig}
\end{figure}

This S/N estimate from repeated observations agrees well with other estimates of the S/N.
Figure \ref{errplot.fig} shows the S/N as a function of wavelength for individual observations of HD 163466, including estimates from observations (solid red squares), bootstrapped from measurements of the RMS of the spectrum about a best-fit spline model (black line), and provided by the ERR array for the data by the JWST calibration pipeline (orange line).  All three estimates agree well over two dex in S/N, and also agree with predictions from the JWST Exposure Time Calculator (ETC; green line).
For a variety of reasons, older versions of the pipeline and associated reference files did not achieve such good agreement in the pipeline-estimated errors (blue line).

\section{Discussion}
\label{discussion.sec}

\subsection{Spectral Discontinuities}

As illustrated in Figure \ref{relphotom.fig}, the MRS spectra often show
percent-level discontinuities at overlapping wavelengths in adjacent spectral bands.  In the bright-source regime these jumps are likely multiplicative and may be due to charge migration, uncertainties in the flatfield \citep[as the DGA wheel does not repeat its position perfectly for each target; see][their Figure 10]{patapis24}, or to uncertainties in our model of the time-dependent loss correction factor.  Such jumps can also occur when using non-standard aperture radii due to imperfections in the aperture correction.
At low flux levels the offsets are likely additive and due to variations in the detector dark current and the challenges of reliably background-subtracting such data.

\subsection{Calibration Trends}

\begin{figure}[!]
\epsscale{1.2}
\plotone{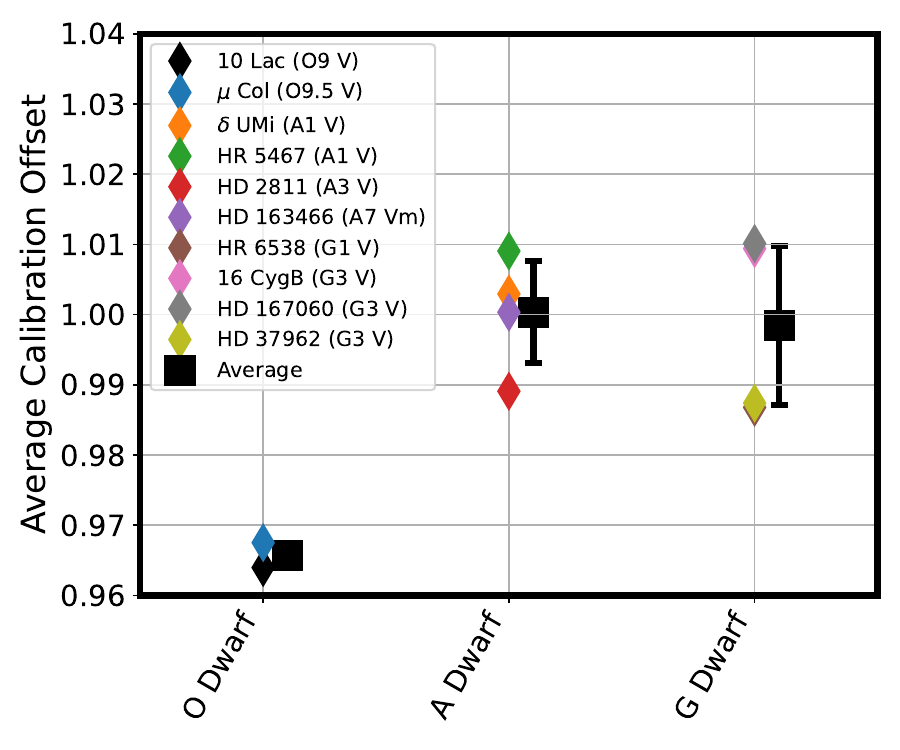}
\caption{Relative calibration factor (averaged across 5--18 $\micron$) derived from each spectrophotometric calibration star as a function of spectral class (colored diamonds).
The average and RMS distribution width within each class are shown as filled squares (offset horizontally for clarity).
}
\label{types.fig}
\end{figure}

Figure \ref{relphotom.fig} shows that
there are systematic differences in the flux calibration vectors derived from different stars.  We average over the wavelength range 5--18 $\micron$ and plot the average offset relative to our final adopted calibration as a function of spectral class in Figure \ref{types.fig}.  
While the A and G dwarfs match the adopted calibration to within an RMS of 1\%, both O stars are consistent with a calibration that is 3\% fainter.  While no such offset was observed for the MIRI imager \citep{gordon24}, with just two such stars observed so far we cannot state with certainty whether this represents a genuine systematic in the CALSPEC models.
Replacing the CALSPEC models with G-star models from \citet{rieke24} results in broadly similar conclusions, although the average G dwarf calibration is then low by a marginally significant $2.4 \pm 0.8$\% compared to the A stars.

\begin{figure*}[!]
\epsscale{1.2}
\plotone{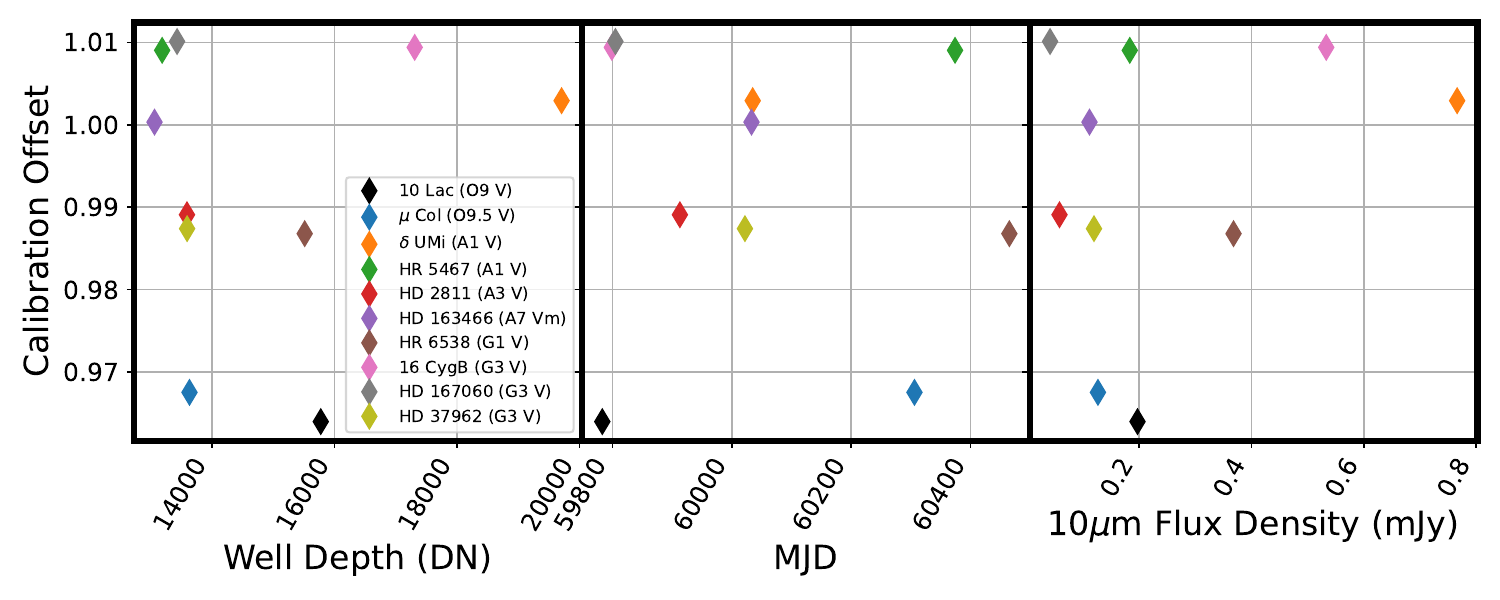}
\caption{Relative calibration factor derived from each spectrophotometric calibration star (colored diamonds) as a function of the maximum detector well depth, modified Julian date (MJD) of observation, and 10 $\micron$ flux density.  The calibration factor does not have a trend with any of these variables, indicating both that the time-dependent count rate loss model is working well and that there are no obvious detector-dependent systematics.
}
\label{instrends.fig}
\end{figure*}

As illustrated by Figure \ref{instrends.fig}, these systematic offsets are unlikely to be
caused by instrumental systematics.
As discussed by \citet{argyriou23b} and \citet{gasman24}, the MRS experiences significant
charge spilling as the total number of recorded counts in a given detector pixel increases,
affecting the non-linearity of both itself and surrounding pixels long before reaching
formal saturation.  However, this effect makes little difference for our observed calibration 
stars (Figure \ref{instrends.fig}, left-hand panel), in part because all reach a similar 
maximum well depth compared to the saturation level of the detector ($\sim$ 55,000 DN).
Indeed, even in cases where the well depth differs significantly, the total extracted flux only changes slightly.  Comparing the calibrated spectra of 10 Lac from PIDs 1524 and 3779, we find that the Channel 1A flux agrees to within 0.5\% despite reaching well depths of 26,000 and 55,000+ DN respectively, and agrees to within 0.2\% in Channel 2A for well depths of 16,000 and 46,000 DN respectively.
Likewise, there is no significant trend of the calibration offsets with the date of observation (Figure \ref{instrends.fig}, middle panel) or the 10$\micron$ flux density of the target star (Figure \ref{instrends.fig}, right panel).

\subsection{Comparison with the MIRI Imager}
\label{imager.sec}

The most direct comparison of our calibration that we can make is against that of the MIRI imager \citep[see][]{dicken24, gordon24}, which observed many of the same standard stars.  We therefore compute the photon-weighted average flux density in units of Jy as

\begin{equation}
    \langle f_{\nu} \rangle = \frac{\int f_{\lambda} \, R_{\lambda} \, \lambda \, d\lambda}{c \int R_{\lambda} \, \lambda^{-1} \, d\lambda}
\label{magnitude.eqn}
\end{equation}

where $f_{\lambda}$ is the MRS flux density (in per-wavelength units), $R_{\lambda}$ is the net photoconversion efficiency (e- per photon) drawn from the JWST Exposure Time Calculator\footnote{https://jwst.etc.stsci.edu/}, and the pivot wavelength is given by

\begin{equation}
    \lambda_{\rm pivot} = \sqrt{\frac{\int R_{\lambda} \, \lambda \, d\lambda}{\int R_{\lambda} \, \lambda^{-1} \, d\lambda}}
\end{equation}

\begin{figure}[!]
\epsscale{1.2}
\plotone{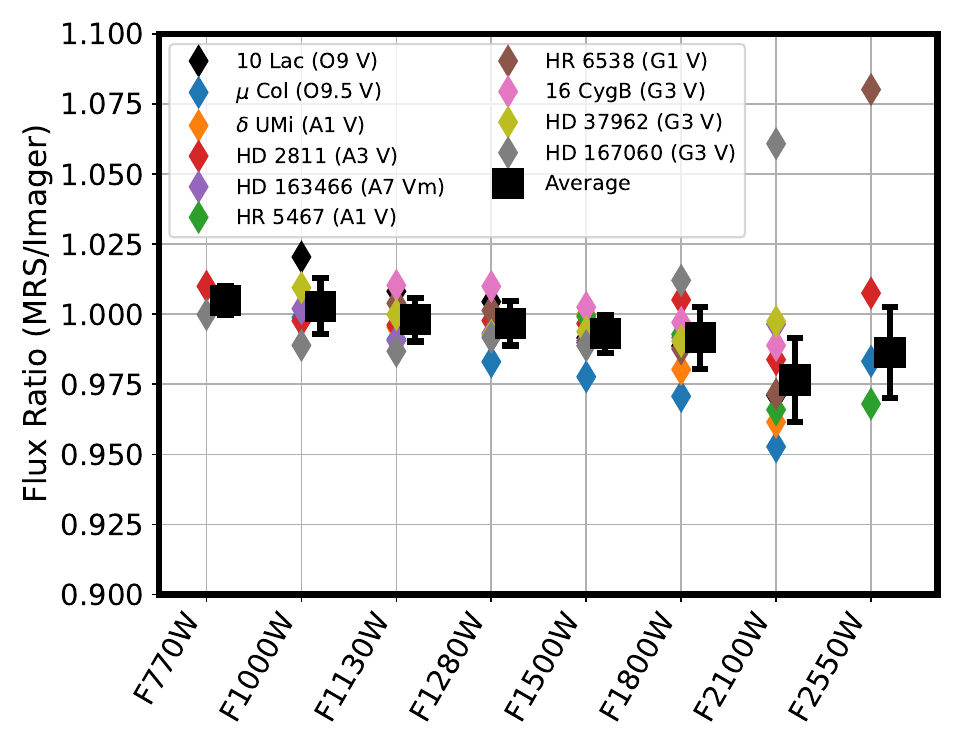}
\caption{Ratio between flux density observed by the MIRI MRS and the MIRI Imager in eight of the nine imaging bandpasses.  Colored diamonds show values from individual stars, while filled squares (offset for clarity) represent the average and RMS distribution within each bandpass.
}
\label{imacompare.fig}
\end{figure}

We divide these results by the values obtained from MIRI imaging observations using the calibration described by \citet{gordon24} and plot the results in Figure \ref{imacompare.fig}.
No comparisons are available in F560W as the MRS calibration targets were too bright to be observed by the imager at this wavelength, while for F2550W we restrict our sample to only those stars with dedicated backgrounds whose readout configuration matches the science data.  This latter step can be important in order to permit pixel-by-pixel subtraction of the dedicated background, minimizing biases that can arise in the annular background subtraction step due to uncertainties in the flatfield correction when sources are extremely faint compared to the thermal background.

In each filter we compute the average calibration offset from the sigma-clipped mean of all individual stars, thereby rejecting two highly discrepant points corresponding to HD 167060 and HR 6538 in F2100W and F2550W respectively.  The MIRI imager and MRS calibrations are consistent with each other to within 1\% from F770W to F1800W, although there is marginal evidence for an offset at the 2.3\% level in F2100W and 1.4\% level in F2550W.
The cause of this offset is uncertain, and may be related to our use of a time-variable source to model the relative response in the Ch4 bandpass.  Additional observations of extremely bright, zeroeth magnitude stars in future calibration programs may help to further characterize this difference.

\subsection{Comparison to Spitzer}

\begin{figure}[!]
\epsscale{1.3}
\plotone{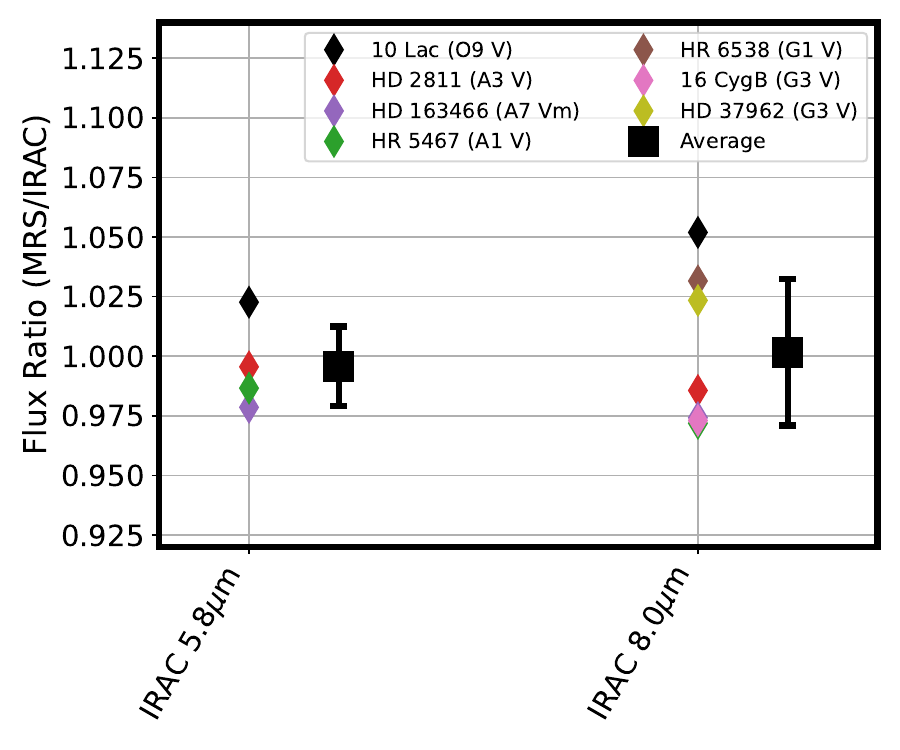}
\caption{Ratio between flux density observed by the MIRI MRS and Spitzer IRAC in the IRAC 5.8 $\micron$ and 8.0 $\micron$ bandpasses.  Colored diamonds show values from individual stars, while filled squares (offset for clarity) represent the average and RMS distribution within each bandpass.
}
\label{irac.fig}
\end{figure}

It is also possible to compare our spectrophotometric calibration of the MIRI MRS against those of the Spitzer Space Telescope Infrared Array Camera (IRAC) and Infrared Spectrograph (IRS) which have been characterized by \citet{reach05} and \citet{sloan15} respectively.
Seven of our calibration stars have been observed with a combination of the IRAC 5.8 $\micron$ and 8.0 $\micron$ filters; these filters overlap sufficiently closely with the wavelength range of the MRS that we can compute synthetic photometry following Eqn \ref{magnitude.eqn} using the filter response functions provided in the IRAC instrument handbook.\footnote{https://irsa.ipac.caltech.edu/data/SPITZER/docs/irac/}
We compare our synthetic magnitudes with the observed IRAC magnitudes tabulated by \citet[][their Table 5]{bohlin22} and note that the average ratio between MRS and IRAC magnitudes is consistent with unity (Figure \ref{irac.fig}), although individual stars can have offsets of up to 5\%.  This average agreement is in large part unsurprising, since the IRAC data have been used to derive the CALSPEC models to which we have tied the MRS calibration.

\begin{figure}[!]
\epsscale{1.3}
\plotone{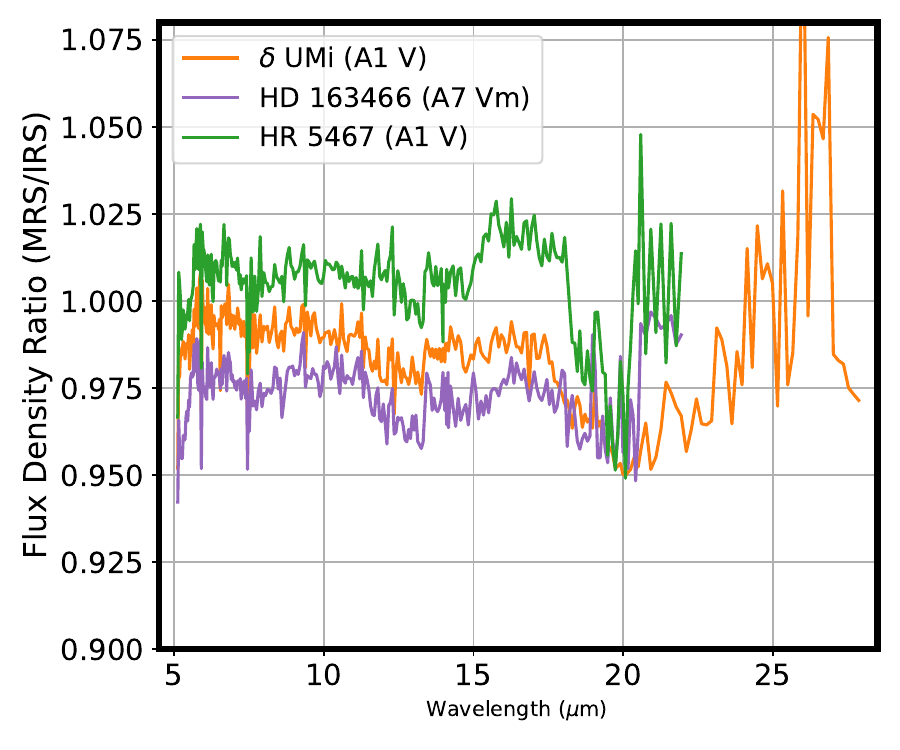}
\caption{Ratio between stellar spectra observed by MIRI MRS and Spitzer IRS (colored lines).  Note that both HD 163466 and HR 5467 have been trimmed at about 20 $\micron$ due to their low SNR at longer wavelengths.
}
\label{spitzer.fig}
\end{figure}

Three A dwarfs were also observed in common by the MIRI MRS and the Spitzer IRS.  Rebinning the MRS spectra to the coarser IRS wavelength scale, Figure \ref{spitzer.fig} plots the ratio of the MRS spectra to the aperture-corrected IRS spectra produced by \citet{sloan15}.
As for the IRAC data, the MRS and IRS data agree to within about 1\% on average below 18$\micron$, although individual stars can each be offset by as much as 3\%.
The calibration for all three stars appear to be systematically offset by a few percent however around 20 $\micron$.  The origin of this offset is uncertain, as it is complicated by background subtraction uncertainties and occurs around the transition wavelength between both MRS Channels 3 to 4 and IRS modules Short-High to Long-High.  The offset is in the same direction as noted in 
\S \ref{imager.sec} for comparisons against the MIRI imager though, suggesting that it may be related to our use of an asteroid calibrator for Channel 4.
These results are broadly consistent with recently-obtained observations of K giants that had previously been used as IRS calibration sources, the comparison with which will be provided in a forthcoming publication (G. Sloan, priv. comm.).

\subsection{Comparison with Giant Planet Observations}

An orthogonal means of evaluating the MRS photometric calibration can be provided using MRS observations of the giant solar system planets Saturn and Uranus.  While our previous absolute flux comparisons have focused on blue point sources (i.e., stars), the giant planets are extended sources (filling the FOV in the case of Saturn), are extremely bright at red wavelengths, and have been studied in detail by the Cassini and Voyager missions.

\begin{figure*}[!]
\epsscale{1.2}
\plotone{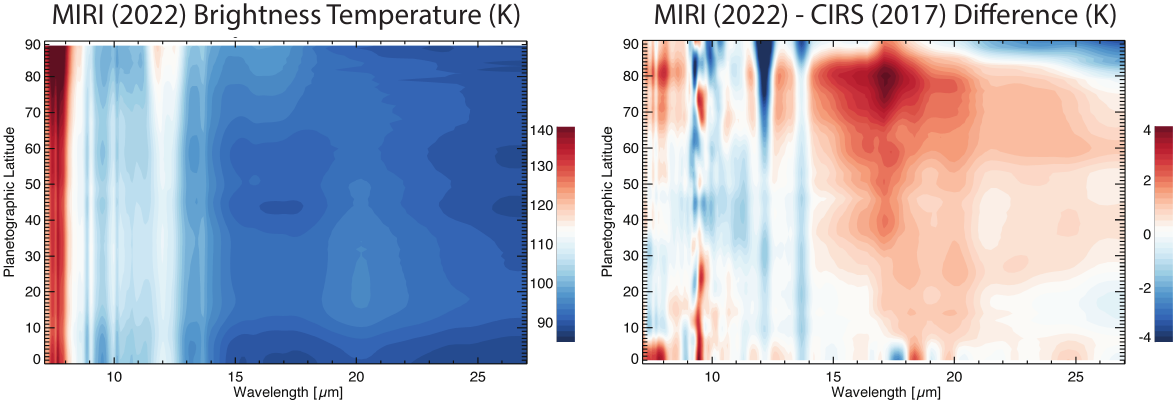}
\caption{Left panel: Observed MIRI MRS brightness temperatures of Saturn in November 2022.  Data were zonally (i.e., east-west) averaged along Saturn's central meridian from the equator to the north pole, and then convolved with the spectral response function of Cassini/CIRS (15 cm$^{-1}$) to display as a function of wavelength.  Right panel: Difference in zonally-averaged brightness temperature between 2022 MIRI MRS observations and 2017 Cassini CIRS observations.
}
\label{saturnbt.fig}
\end{figure*}

In Figure \ref{saturnbt.fig} we show a brightness temperature map of Saturn constructed using our adopted flux calibration and following the methodology of \citet{fletcher23}.  Spectra from three MIRI/MRS footprints spanned the equator to Saturn’s north pole, and were zonally (i.e., east-west) averaged to display as a function of latitude and wavelength.  These data were acquired in November 2022 (program ID 1247) during mid-northern-summer on Saturn, five years after the end of the Cassini spacecraft exploration of Saturn (2017, northern summer solstice).  To provide a fair comparison to Cassini data, we convolved the MRS data with the 15 cm$^{-1}$ spectral response function of the Composite Infrared Spectrometer (CIRS).  CIRS maps acquired during Cassini’s final year were also averaged, and the difference is shown in Figure \ref{saturnbt.fig}.  

Despite the 5 years of elapsed time between the measurements (i.e., 1/6th of a Saturnian year), the differences shown in Figure \ref{saturnbt.fig} are entirely within expectations for this seasonal giant planet.  This region of the spectrum is dominated by collision-induced opacity of hydrogen and helium, so can be used as a reliable thermometer for tropospheric temperatures.  The MIRI spectra are about 1 K warmer on average in the 22-28 $\micron$ range (sounding the troposphere down to ~1 bar, cf contribution functions calculated in Fig. 13 of \citet{fletcher23}), and about 3-4 K warmer in the 17 $\micron$ range (sounding the upper troposphere and lower stratosphere).  As seasonal influences are more extreme at higher altitudes and lower pressures, this demonstrates that Saturn’s summertime warming has been faster at the tropopause than at deeper atmospheric levels.  

Converting from brightness temperature to flux density, these differences correspond to about 6\% and 30\% increases in flux density respectively.  The largest differences occur at Saturn’s high latitudes, where a polar vortex is known to exhibit large seasonal changes in atmospheric temperature.  These differences are entirely within expectation for natural seasonal variations in Saturn’s atmosphere 
\citep[e.g.,][]{fletcher20}, but the variability of the giant planet prevents a more quantative assessment of the MRS flux uncertainty.   Finally, although ground-based observations of Saturn are available during the same period, these suffer from significant calibration uncertainties and cannot aid this comparison.
 
However, a similar exercise can be performed for a giant planet where expected seasonal variability is even smaller.  Uranus was explored in detail by Voyager 2 in 1986, when the planet’s southern pole faced towards Earth (southern summer solstice).  The Infrared Interferometer Spectrometer and Radiometer (IRIS) mapped tropospheric temperatures using far-infrared wavelengths from 25-50 µm.  As discussed by M. Roman et al. (2024, in prep), MIRI/MRS observed Uranus during mid-northern spring in 2023, and the full 5-28 µm range was used to derive temperatures as a function of latitude and pressure.  Using our adopted flux calibration, tropospheric temperatures inferred from both spacecraft – despite covering different spectral ranges, with different instrumentation, and nearly four decades apart – match one another within $\sim$0.4 K.  At a typical temperature of 55 K (appropriate for the 250-mbar level), and assuming no unexpected evolution in the atmosphere of Uranus, this match thus implies that the MIRI Channel 4 flux calibration is accurate to within 5\%.

\section{Summary}

We have presented the default JWST pipeline spectrophotometric calibration of the MIRI MRS.  As key preliminary steps we derived detector pixel flatfields using self-calibration techniques applied to observations of the planetary nebula NGC 7027 and aperture corrections from a combination of PSF models and observations of bright stellar targets.  We discussed evidence for a strong time-dependent decrease in the effective count rate that grows in strength with wavelength, and constructed a model of this loss that corrects all MRS observations back to a nominal zero-day count rate.

Using these time-corrected count rates, we used deep observations of the O9 V
standard star 10 Lac in combination with CALSPEC spectral models to derive 
an initial set of photometric response vectors for Channels 1A--3C (i.e., shortward of 18$\micron$).  We then adjusted the normalization of this model by $\sim$ 4\%
to match the overall average of 10 different calibration
stars including O, A, and G dwarfs.  In Channel 4 (longward of 18 $\micron$) we used observations of the main-belt asteroid 515 Athalia and young stellar object SAO 206462 to derive the photometric response based on their smooth, predictable spectra.

This calibration is repeatable over time to sub-percent accuracy in Channels 1-3, with no evidence for systematic offsets with detector well depth or stellar flux density.  Likewise, the point-source calibration matches that of the MIRI imager to within 1\% and prior Spitzer IRS observations to within 1--3\%.  The MRS extended-source flux calibration depends upon the accuracy of the PSF model and aperture corrections, but comparisons against both Voyager and Cassini observations of giant planets suggests that it is accurate to within $\sim$ 5--6\%.

The calibrations described in this paper are available in CRDS context 1276.  The best calibrations used by the JWST pipeline will, however, continue to evolve over time as additional data are obtained and the characterization of the MRS instrumental performance improves.

\begin{acknowledgments}

This work is based on observations made with the NASA/ESA/CSA James Webb Space Telescope. The data were obtained from the Mikulski Archive for Space Telescopes at the Space Telescope Science Institute, which is operated by the Association of Universities for Research in Astronomy, Inc., under NASA contract NAS 5-03127 for JWST. These observations are associated with programs \# 1050, 1282, 1518, 1523, 1524, 1536, 1538, 1539, 1549, 3779, 4484, 4489, 4496, 4497, 4498, 4499, 6607, and 6612.

The following National and International Funding Agencies funded and supported the MIRI development: NASA; ESA; Belgian Science Policy Office (BELSPO); Centre Nationale d'Etudes Spatiales (CNES); Danish National Space Centre; Deutsches Zentrum fur Luftund Raumfahrt (DLR); Enterprise Ireland; Ministerio De Economi´a y Competividad; Netherlands Research School for Astronomy (NOVA); Netherlands Organisation for Scientific Research (NWO); Science and Technology Facilities Council; Swiss Space Office; Swedish National Space Agency; and UK Space Agency.

D. Law appreciates productive discussions and input from Michael Roman and Oliver King.
L. Fletcher was supported by STFC Consolidated Grant reference ST/W00089X/1

\end{acknowledgments}

\appendix
\section{Standard Star Spectra}
\label{appendix.sec}

\begin{figure*}[!]
\epsscale{1.1}
\plotone{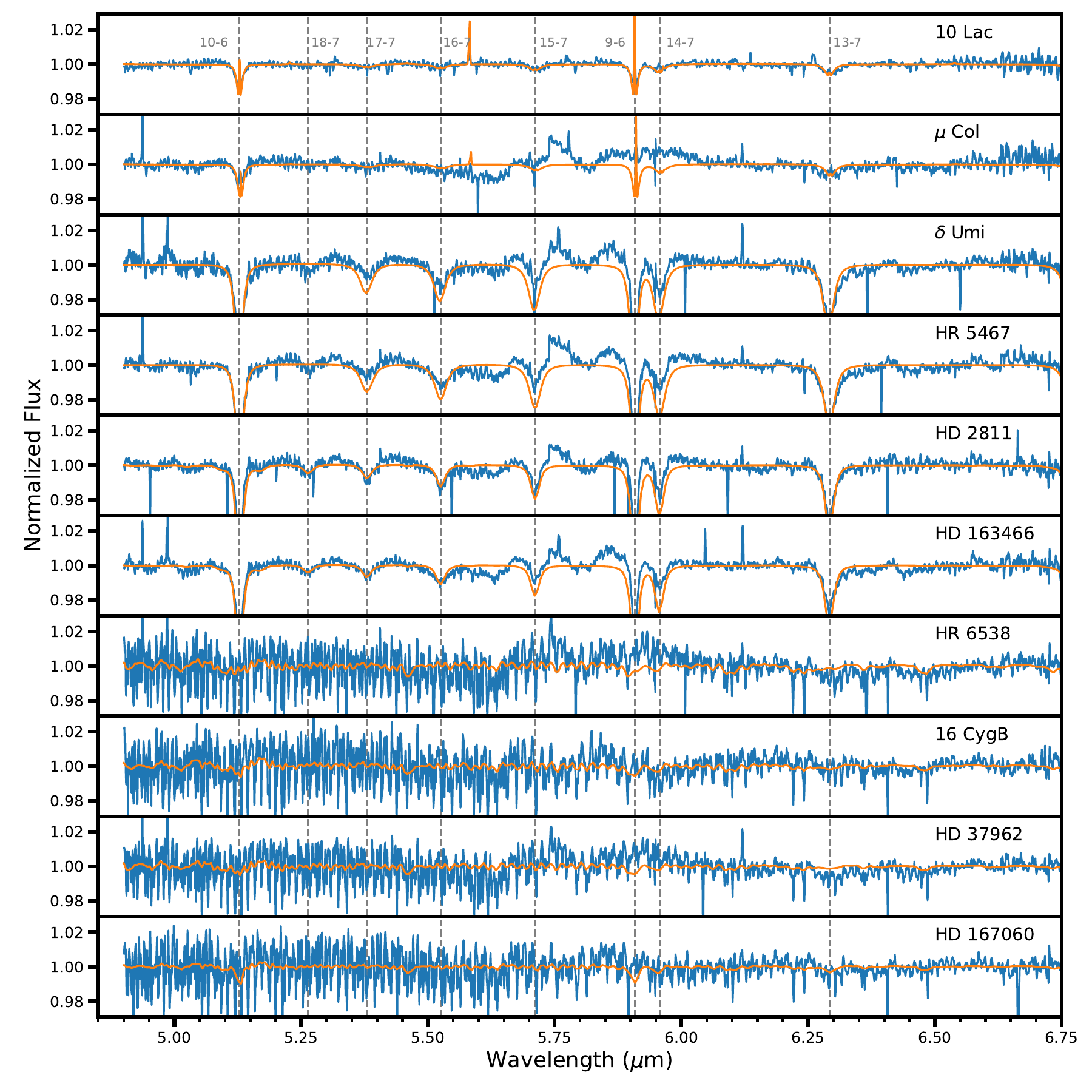}
\caption{Continuum-normalized spectra (solid blue lines) and normalized CALSPEC models (solid orange lines) for each of the ten observed standard stars in the wavelength range $\lambda = 4.9-6.7$ \micron.  Dashed vertical lines in all panels represent the wavelengths of major atomic hydrogen transitions.
}
\label{spectra_part1.fig}
\end{figure*}

\begin{figure*}[!]
\epsscale{1.1}
\plotone{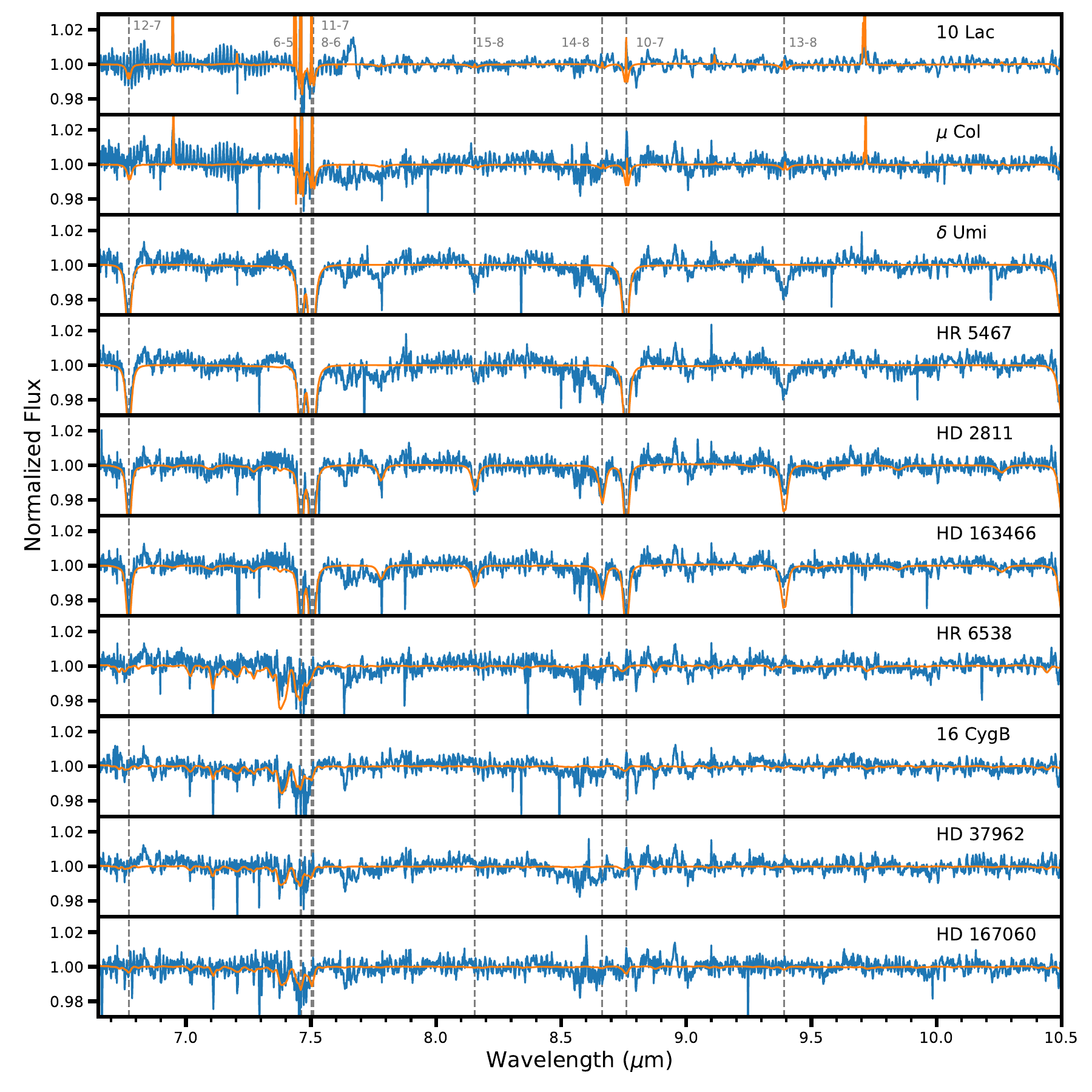}
\caption{As Figure \ref{spectra_part1.fig} but for the wavelength range $\lambda = 6.7-10.5$ \micron.
}
\label{spectra_part2.fig}
\end{figure*}

In Figures \ref{spectra_part1.fig} - \ref{spectra_part2.fig} we plot the continuum normalized spectra of each of the ten standard stars used in our flux calibration analysis, along with the corresponding normalized CALSPEC model.  Any large-scale multiplicative offsets between observations and the models have been removed by this continuum normalization, and these plots are thus most useful for identifying unexpected spectral features.  We note in particular broad high-ionization features in 10 Lac \citep[see \S \ref{nev.sec} and][]{law24} and high-frequency structure below 6$\micron$ in the spectra of the four G-dwarf stars that may be due to CO band features.

\begin{figure*}[!]
\epsscale{1.1}
\plotone{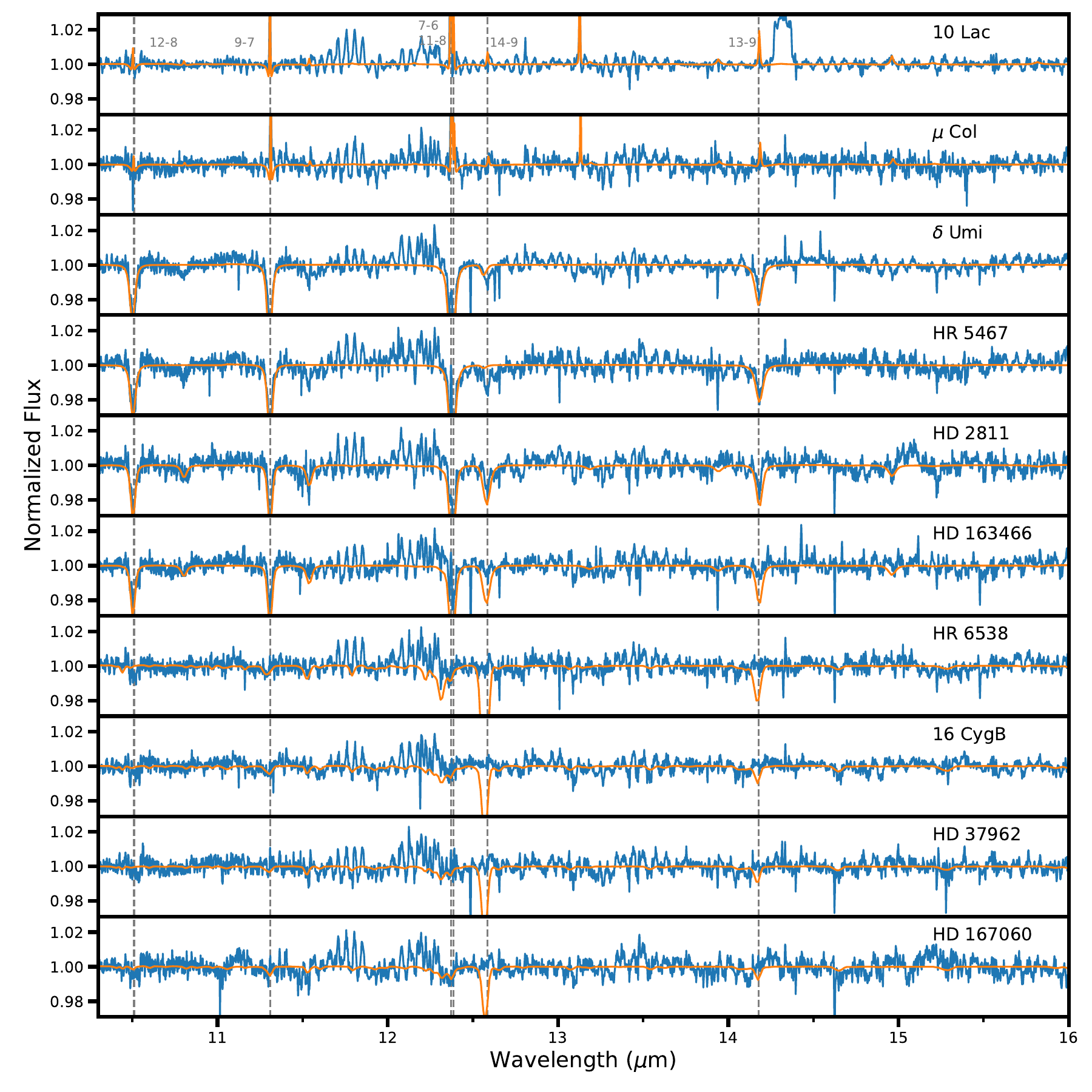}
\caption{As Figure \ref{spectra_part1.fig} but for the wavelength range $\lambda = 10.5-16.0$ \micron.
}
\label{spectra_part3.fig}
\end{figure*}

\begin{figure*}[!]
\epsscale{1.1}
\plotone{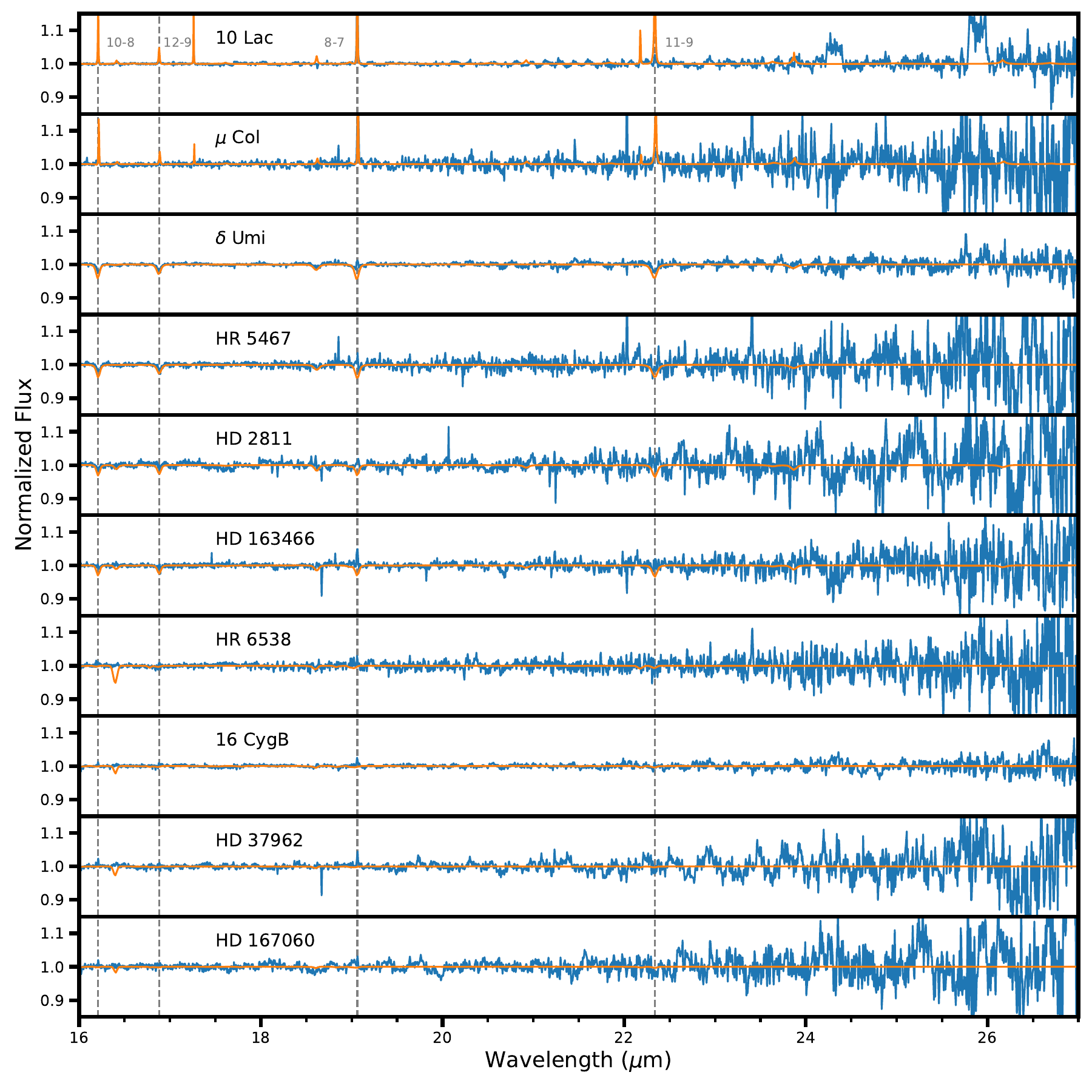}
\caption{As Figure \ref{spectra_part1.fig} but for the wavelength range $\lambda = 16.0-27.0$ \micron.  Note that the vertical scale is much larger than in Figures \ref{spectra_part1.fig} - \ref{spectra_part3.fig} to accommodate the lower SNR at long wavelengths.
}
\label{spectra_part4.fig}
\end{figure*}

\end{document}